\documentclass[aps,prl,reprint,superscriptaddress,floatfix]{revtex4-2} 
\usepackage{amsmath}
\usepackage{amssymb}
\usepackage{mathtools}
\usepackage{graphicx} 
\usepackage{xcolor}
\usepackage{url}
\usepackage{comment}

\usepackage{hyperref}
\hypersetup{colorlinks=true,urlcolor=blue,linkcolor=blue,citecolor=blue}
\usepackage{orcidlink} 

\DeclareMathOperator{\Tr}{Tr}
\newcommand{\Hint}{H_{\text{int}}}
\newcommand{\beff}{\beta_{\mathrm{eff}}}
\newcommand{\Gb}{\bar{G}}

\DeclarePairedDelimiter\bra{\langle}{\rvert}
\DeclarePairedDelimiter\ket{\lvert}{\rangle}
\DeclarePairedDelimiterX\braket[2]{\langle}{\rangle}{#1\,\delimsize\vert\,\mathopen{}#2}

\begin{document}
\bibliographystyle{apsrev4-2}
\title{Size Operator and Spectral Clustering in the Two Coupled SYK Model}

\author{Juan Santos-Suárez\orcidlink{0000-0001-9360-2411}}
\email{juansantos.suarez@usc.es}
\author{Martí Berenguer\orcidlink{0000-0002-3791-9585}}
\email{marti.berenguer.mimo@usc.es}
\author{Javier Mas\orcidlink{0000-0001-7008-2126}}
\email{javier.mas@usc.es}
\author{Alfonso V. Ramallo\orcidlink{0000-0003-2762-2873}}
\email{alfonso.ramallo@usc.es}

\affiliation{Departamento de Física de Partículas, Universidade de Santiago de Compostela and Instituto Galego de Física de Altas Enerxías (IGFAE), 15782 Santiago de Compostela, Spain}

\begin{abstract}
At large $N$, two coupled Sachdev--Ye--Kitaev models realize an eternal traversable wormhole with a discrete spectrum. We show that at finite $N$ the spectrum organizes into clusters labeled by operator size. Low-size clusters evolve into the conformal towers and define a weak-ergodicity-breaking subspace responsible for the long-lived wormhole revival dynamics. Moreover, the competition between size energy and size entropy provides a microscopic interpretation of the wormhole--black hole transition. These results reveal operator size as the bridge between the finite-$N$ spectrum and the emergent gravitational physics at large $N$.
\end{abstract}

\maketitle

\emph{ \color{blue}Introduction.--} Quantum systems with a holographic gravity dual are coming within experimental reach of quantum simulators~\cite{Joshi2022, Uhrich2023,Baumgartner2024,Creffield2026, Biswas2026, Steiner2026, Bang2026}. However, the gravitational picture is best understood in the large-$N$ limit, while quantum simulators can only realize finite-$N$ systems. On a quantum processor, a momentary coupling between two chaotic systems was used to teleport a signal through what was reported as a traversable wormhole~\cite{Gao2017,Gao2021,Jafferis2022,Schuster2022,Brown2023,Nezami2023}. A debate followed about which features of a finite system can be interpreted as gravitational~\cite{Kobrin2023,Jafferis2023}.

An eternal traversable wormhole in nearly-AdS$_2$ gravity is realized instead when the coupling is left permanently on. Its microscopic dual is the large-$N$ low-temperature limit of the two coupled Sachdev--Ye--Kitaev (SYK) model~\cite{MQ}, which joins two maximally chaotic systems~\cite{Sachdev_1993,KitaevTalk1,KitaevTalk2,MaldaStanford,Polchinski2016} through a simple bilinear term. The emergent nearly-conformal symmetry of that regime organizes the low-energy excitations into a matter tower and a graviton tower whose spacing is set by an infrared scale~\cite{MQ,Lantagne2020}. This discrete spectrum produces the wormhole's defining dynamical signature, the long-lived periodic revivals of signals transmitted between the two boundaries~\cite{Lantagne2020}. The structure survives only at low temperatures. Upon heating, a phase transition to a two-black-hole geometry~\cite{MQ,Lantagne2020,FloquetWormhole,HotWormholeExponents,Zhang2026} dissolves the towers and, with them, the revivals.

That tower structure is the spectral hallmark of the wormhole description. Deciding which features of the finite-$N$ system may be attributed to a gravitational dual therefore comes down to a sharper question: what becomes of the towers at finite $N$, where they must be built from individual eigenstates? In critical systems such an emergent conformal spectrum is directly visible in the microscopic one, with excitation energies fixed by the scaling dimensions through the operator--state correspondence~\cite{Cardy1984,Cardy1986,Bloete1986,Affleck1986,Milsted2017,Zou2018,Sun2026}. Nothing guarantees the same here, since chaotic Hamiltonians are generically expected to follow random matrix theory rather than to display sharp spectral ladders~\cite{Haake2018}. Yet wormhole physics at finite $N$ is being studied through exact diagonalization~\cite{GarciaGarcia2019,Alet2021,Caceres2021,HotWormholeExponents} and quantum simulation~\cite{Brown2023,Nezami2023,Schuster2025}. Such a program requires understanding what a conformal tower level is at finite $N$, and which quantum number labels it.

In this Letter we show that the finite-$N$ spectrum of the coupled model is organized into clusters labeled by operator size~\cite{Roberts2018,QiStreicher2019,Zhang2023,Schuster2022}. An exact $\mathbb{Z}_4$ symmetry~\cite{GarciaGarcia2019} restricts the mixing between size sectors so that the cluster width shrinks with $N$. The conformal towers then arise not from individual eigenstates but from well-resolved clusters. Dynamically, these clusters evolve coherently as rigid units. Their nearly uniform spacing produces the periodic transmission revivals and their finite internal widths sets the decoherence time through dephasing. This coherent dynamics defines an emergent weak-ergodicity-breaking subspace, placing the wormhole physics within the framework of quantum many-body scars~\cite{Turner18,Sala2020,Ren21,Papic2022,Yu2025,Milekhin2024}. The competition between the size energy and size entropy of the clusters is the microscopic precursor of the wormhole--black hole transition. The conformal towers, the revival dynamics and the phase transition thus emerge as different manifestations of a single underlying spectral organization.

\emph{ \color{blue}The model.--} The two coupled SYK model~\cite{MQ} consists of two identical SYK models, labeled Left ($L$) and Right ($R$). Each side comprises $N$ Majorana fermions $\chi_j^a$ with $a \in \{L, R\}$ satisfying $\left\{\chi_i^a,\chi_j^b\right\}=\delta_{ij}\delta^{ab}$. The total Hamiltonian is defined as 
\begin{equation}
    H(\mu)= H_{\text{SYK}}^L + H_{\text{SYK}}^R + \mu \Hint,
\end{equation}
where both sides carry the same SYK Hamiltonian
\begin{equation}
    H_{\text{SYK}}^a = i^{q/2}\!\!\sum_{j_{1}<\cdots<j_{q}}\!\!J_{j_{1}\cdots j_{q}}\chi_{j_{1}}^a\cdots\chi^a_{j_{q}}, \qquad a=L,R,
\end{equation}
with zero-mean Gaussian random couplings of variance $\langle J^2\rangle = (q-1)!\,J^2/N^{q-1}$. Throughout this work we set $q=4$ and $J=1$. The left--right coupling is
\begin{equation}
    \Hint = i\sum_{j=1}^{N}\chi_{j}^{L}\chi_{j}^{R}.
\end{equation}
This interaction term diagonalizes in the basis $d_j = (\chi_j^L+i\chi_j^R)/\sqrt{2}$, where $\Hint = Q - N/2$ with $Q=\sum_j d_j^\dagger d_j$~\cite{GarciaGarcia2019}. Its spectrum is that of $N$ decoupled fermionic oscillators $Q\ket{k,m}=k\ket{k,m}$, with $k=0,1,\ldots,N$ and $m=1,\ldots,N_k\equiv\binom{N}{k}$ accounting for the binomial degeneracies. The projector to a sector $\mathcal{H}_k$ with eigenvalue $k$ is thus given by $P_k=\sum_{m=1}^{N_k}\ket{k,m}\bra{k,m}$. The ground state of $Q$ corresponds to the infinite temperature Thermofield Double (TFD) state, $\ket{0,0}\equiv\ket{\mathrm{TFD}(\beta=0)}$~\cite{GarciaGarcia2019}. Under the operator-state mapping  $Q$ coincides with the Qi--Streicher size operator~\cite{Roberts2018,QiStreicher2019,Zhang2023, Schuster2022} whose eigenvalues $k$ count the number of Majoranas in each basis string, so its expectation value defines the average operator size. 

\begin{figure*}[!t]
    \centering
    \includegraphics[width=\linewidth]{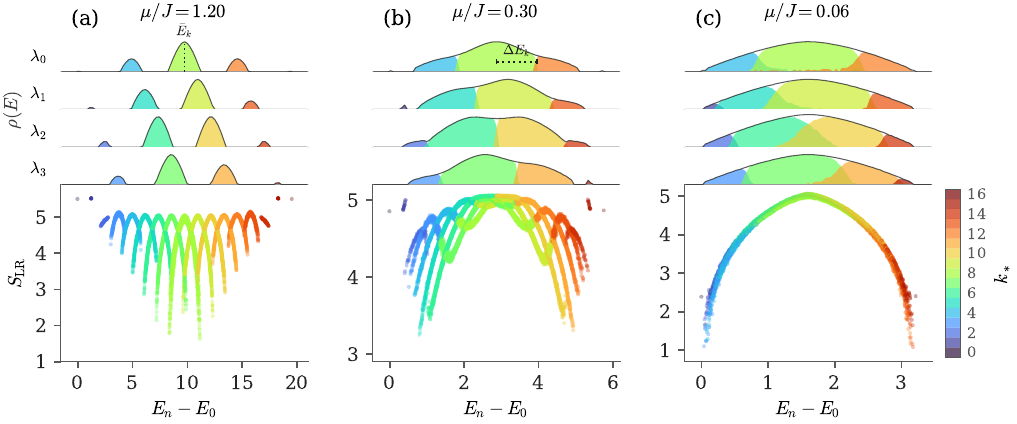}
    \caption{Spectral clustering for $N=16$, $q=4$ for $\mu/J=1.20,\,0.30,\,0.06$ (a--c) (one disorder realization; no significant variation between realizations is observed). Bottom row: left--right entanglement entropy $S_{\mathrm{LR}}$ of every eigenstate versus excitation energy $E_n-E_0$, colored by dominant size $k_*$. Top rows: the density of states $\rho(E)$ resolved by $\mathbb{Z}_4$ symmetry sector $\lambda_r$. The faint black line in each row is the total sector density while the colored fills centered at $\bar E_k$ (marked in (a)) with width $\Delta E_k$ (marked in (b)) show its decomposition into individual sizes $k_*$. The sizes appear as well-separated peaks in (a), broaden and overlap in (b), and mix strongly in (c).}
    \label{fig:clustering}
\end{figure*}

A discrete $\mathbb{Z}_4$ symmetry generated by $e^{i\pi \Hint/2}$, with eigenvalues $\lambda_r = i^{-N/2}i^r$ for $r \in\{0,1,2,3\}$, allows one to block-diagonalize $H(\mu)$ into four sectors of dimension $\sim 2^{N-2}$~\cite{GarciaGarcia2019,SM}.

In the solvable large-$N$ limit, the low-energy spectrum of the \emph{traversable wormhole} phase organizes into a \emph{matter} tower and a \emph{graviton} (boundary-reparametrization) tower~\cite{MQ}. Their energy levels are fixed by an emergent conformal scale $\varepsilon\sim J(\mu/J)^{\frac{q}{2(q-1)}}=\mu^{2/3}$,
\begin{equation}
E_n^{(m)}=E_{\rm gap}^{(m)}\,(qn+1),\qquad E_n^{(g)}=E_{\rm gap}^{(g)}\,(2n+1),
\label{eq:towers_largeN}
\end{equation}
with $E_{\rm gap}^{(m)}=\varepsilon/q$ and $E_{\rm gap}^{(g)}=\varepsilon\sqrt{(q-1)/2q}$ so that $E_{\rm gap}^{(m)}<E_{\rm gap}^{(g)} = \mathcal{O}(\varepsilon)$~\cite{MQ,Lantagne2020}. These ratios hold asymptotically as $\mu\to 0$. At finite coupling the large-$N$ spectrum obtained from the Schwinger--Dyson (SD) equations interpolates smoothly between the harmonic ladder at $\mu\gg J$ and the conformal towers~\cite{MQ,Lantagne2020,FloquetWormhole}. This discrete spectrum, governed by the scale $\varepsilon$, is the hallmark of the nearly-AdS$_2$ regime and its wormhole physics. We now show that the microscopic objects that form these towers are not individual eigenstates but operator-size clusters.

\emph{\color{blue} Operator size and emergent clustering.--}
The role of operator size is most transparent in the harmonic limit $\mu\gg J$ where $H(\mu) \approx \mu \Hint$. The spectrum consists of $N+1$ degenerate energy levels labeled by the size $k$ and separated by the harmonic spacing $\mu$. Away from this limit, although $Q$ no longer commutes with $H(\mu)$, the $\mathbb{Z}_4$ symmetry restricts the mixing to $\Delta k\equiv0\ (\mathrm{mod}\ 4)$, while the $q=4$ nature of the SYK terms bounds its reach to $|\Delta k|\le 4$. The Hamiltonian is thus banded in the size label $k$~\cite{GarciaGarcia2019, Kota2001}. This structural constraint protects the different size sectors from complete hybridization. As we now show, despite the lack of exact size conservation, the low-size sectors remain spectrally resolved and organize the finite-$N$ spectrum into distinct operator-size clusters. 

To characterize these clusters quantitatively, operator size must first be extended away from the harmonic limit. Size is meaningful only relative to a reference state and at finite $\mu$ the ground state of $H(\mu)$ is no longer the Fock state $\ket{\rm TFD(\beta=0)}$ annihilated by the $d_j$. We therefore label each eigenstate of $H(\mu)$ by adiabatic continuation from the Fock space at $\mu \to \infty$, and define the size sector $\mathcal{H}_k(\mu)$ as the span of the eigenstates carrying the original label $k$. Each sector occupies a finite energy window of spectral radius $\Delta E_k$ about its centroid $\bar E_k$. 

A perturbative analysis in the $\mu\gg J$ limit extended via a Feshbach--Fano partitioning~\cite{Feshbach1964,Fano1961,Lowdin1962} to all $\mu/J$ reveals that this broadening is constrained to $\Delta E_k\sim k/\sqrt N$ for $k>1$, while for $k=1$ it is further suppressed as $\Delta E_1\sim 1/N$~\cite{SM}. The same partitioning shows that the centroids get displaced, replacing the harmonic scaling $\mu$ by a renormalized scale that, as we will numerically show below, is consistent with the conformal scale $\varepsilon\sim\mu^{2/3}$. A sector therefore remains spectrally resolved as a distinct \emph{cluster} whenever its width $\Delta E_k$ stays below the spacing to its neighbor in the same symmetry sector, $\Delta_k=\bar E_{k+4}-\bar E_k$~\cite{Zelevinsky1996}. Since the widths shrink with $N$ whereas the spacing $\Delta_k$ is nearly uniform at low $k$, $\Delta_k\simeq\Delta_0$, clusters with $k<k_c\sim\sqrt{N}\,\Delta_0/J$ become progressively better resolved as $N$ increases and stay resolved down to smaller $\mu/J$~\cite{SM}. Thus, in the large-$N$ limit low-size clusters turn into degenerate eigenspaces of $H(\mu)$ for all $\mu$, as in the quasi-symmetry construction of Ref.~\cite{Ren21}, but they are not eigenspaces of $Q$.

Because building the spaces $\mathcal{H}_k(\mu)$ requires adiabatically tracking operators of unbounded fermionic weight, we approximate the adiabatic flow using the near-thermal nature of the ground state, $\ket{E_0(\mu)}\approx\ket{\mathrm{TFD}(\beff(\mu))}$~\cite{MQ,Cottrell2019,Alet2021,Caceres2021,GarciaGarcia2019,Schuster2025,Khor2026}. The non-unitary similarity transformation $S(\mu)=e^{-\beff(\mu)(H^L_{\text{SYK}}+H^R_{\text{SYK}})/4}$ maps the Fock vacuum onto this TFD, namely, $\ket{\mathrm{TFD}(\beff)}\propto S\ket{\mathrm{TFD}(\beta=0)}$. This allows us to define a dressed size operator $\tilde Q(\mu)=SQS^{-1}$ that counts size excitations above the approximate ground state~\cite{QiStreicher2019,Roberts2018}. $\tilde Q(\mu)$ remains isospectral to $Q$, and its oblique projectors $\tilde P_k(\mu)=SP_kS^{-1}$ serve as controlled estimators for the exact projectors obtained using the adiabatic flow~\cite{SM}. The distribution of a sector over the spectrum is then captured by the real-valued Kirkwood--Dirac quasiprobability $\mathcal{W}_k(n)=\bra{E_n}\tilde P_k(\mu)\ket{E_n}$~\cite{Hofmann2012,YungerHalpern2018}.

\begin{figure}
    \centering
    \includegraphics[width=\linewidth]{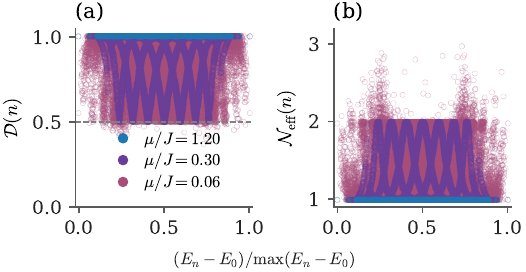}
    \caption{Clustering diagnostics of eigenstates of Fig.~\ref{fig:clustering}. (a) Cluster dominance $\mathcal{D}(n)$ and (b) effective cluster number $\mathcal{N}_{\rm eff}(n)$. See text for interpretation.}
    \label{fig:diagnostics}
\end{figure}

We work with its positive normalized truncation $W_k(n)=\max(0,\mathcal{W}_k(n))/\sum_{k'}\max(0,\mathcal{W}_{k'}(n))$, with $\tilde N_k=\sum_nW_k(n)\approx N_k$. Size averages are $\langle\cdots\rangle_k\equiv\tilde N_k^{-1}\sum_nW_k(n)(\cdots)$, fixing the cluster centroid to $\bar E_k=\langle E_n\rangle_k$. Each eigenstate carries a dominant size $k_*(n)=\operatorname{argmax}_kW_k(n)$ together with a dominance $\mathcal{D}(n)=W_{k_*}(n)$ and an effective cluster number $\mathcal{N}_{\text{eff}}(n)=(\sum_kW_k(n)^2)^{-1}$ that measure its dispersion across the dressed size sectors~\cite{SM}.

Exact diagonalization reveals the cluster organization anticipated above. At $\mu/J=1.20$ the density of states of each $\mathbb{Z}_4$ sector shows well-separated maxima and the left--right entanglement entropy $S_{\mathrm{LR}}$ resolves into the corresponding branches (Fig.~\ref{fig:clustering}(a)). The size assignment is essentially binary, $\mathcal{D}\simeq1$ and $\mathcal{N}_{\text{eff}}\simeq1$ (Fig.~\ref{fig:diagnostics}). Upon reducing the coupling to $\mu/J=0.30$, the clusters in the bulk of the spectrum begin to overlap while those at the edges remain well resolved; most eigenstates are still concentrated in a single size sector (Fig.~\ref{fig:clustering}(b)). By $\mu/J=0.06$ the widths become comparable to $\Delta_k$ throughout the spectrum, the entropy profile approaches that expected from RMT~\cite{Guhr1998, Rodgers1988,Papic2022} (Fig.~\ref{fig:clustering}(c)), and the size ceases to be resolved in energy. 

\begin{figure*}
    \centering
    \includegraphics[width=\linewidth]{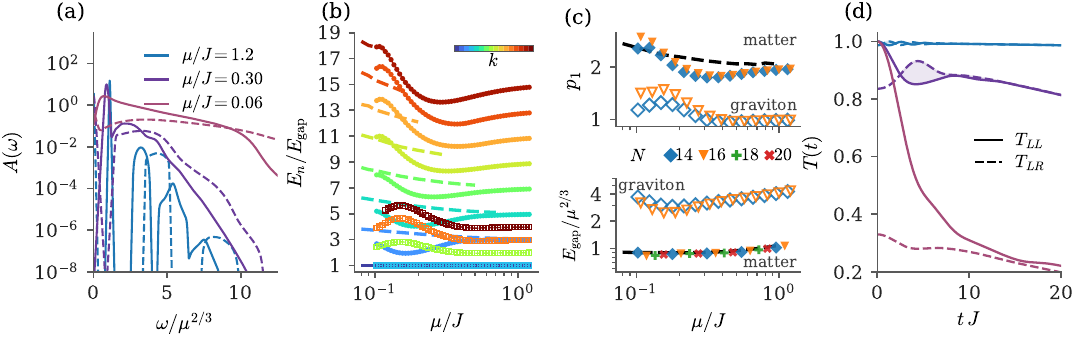}
    \caption{Spectral functions, transmission revivals and tower positions for $N=16$, $q=4$ averaged over $50$ disorder realizations. (a) Zero-temperature matter and graviton spectral functions $A^{(m)}(\omega)$ (solid), $A^{(g)}(\omega)$ (dashed). The $\delta$-peaks in Eq.~\eqref{eq:spectral_functions} are smeared by a normalized kernel of width $\eta$, chosen larger than the mean level spacing so that individual states are not resolved. (b) Ratios of the peak positions to the first peak in both towers, colored by cluster size $k$ (colorbar). Here we evaluate the cluster centroids only over the states unambiguously assigned to a single cluster ($\mathcal{D}(n)>0.7$). Thick dashed lines are the large-$N$ SD reference for the matter tower. (c) Fits of these ratios to $E_n/E_{\rm gap} = p_1n+p_0$, shown in two stacked sub-panels: slope $p_1$ and gap $E_{\rm gap}/\mu^{2/3}$ (the intercept $p_0\simeq1$ throughout and is not shown); marker and color label the system size $N=14$--$20$, and the black dashed curve is the large-$N$ SD reference for the matter tower. (d) Transmission amplitudes $T_{LL}(t)$ (solid) and $T_{LR}(t)$ (dashed), exhibiting alternating revivals when clusters are resolved. Colors in (a) and (d) label the coupling $\mu/J$.}
    \label{fig:towers}
\end{figure*}

\emph{ \color{blue}Conformal towers from clusters.--} At large $N$, the holographic towers of Eq.~\eqref{eq:towers_largeN} appear as poles of zero-temperature spectral functions of a probe $O$~\cite{Lantagne2020,FloquetWormhole},
\begin{equation}
A^{O}(\omega)=\frac{1}{N}\sum_{j,n}\big|\bra{E_n}O_j\ket{E_0}\big|^{2}\, \delta\big(\omega-(E_n-E_0)\big), \label{eq:spectral_functions}
\end{equation}
with $O_j^{(m)}=\chi^L_j$ for the matter tower and $O_j^{(g)}=i\chi^L_j\chi^R_j$ for the graviton one~\cite{Schuster2025}. Every eigenstate contributes a $\delta$-peak, so whenever the clusters are resolved each spectral function breaks into a discrete sequence of resonances, one per cluster, riding on a smoothly decaying envelope [Fig.~\ref{fig:towers}(a)]. The $\mathbb{Z}_4$ symmetry fixes which clusters each probe can reach. Chirality restricts the matter probe to the sectors $\lambda_{1,3}=\pm i\lambda_{0}$, that is, odd $k$, while the graviton probe stays in the ground-state sector $\lambda_{0}$ and sees only $k\equiv 0\ (\mathrm{mod}\ 4)$~\cite{SM}. 

Therefore, we identify the clusters as the finite-$N$ precursors of the towers, $E^{(m)}_{n}=\bar E_{2n+1}-E_0$ and $E^{(g)}_{n}=\bar E_{4(n+1)}-E_0$. The resonance positions follow the linear law $E^{(m),(g)}_{n}\simeq E^{(m),(g)}_{\rm gap}(p_1^{(m),(g)}n+p_0^{(m),(g)})$ [Fig.~\ref{fig:towers}(b,c)]. At large $\mu/J$ the ladders are harmonic, $E^{(m)}_{n}\sim\mu(2n+1)$ and $E^{(g)}_{n}\sim4\mu(n+1)$, i.e., $(p_1^{(m)},p_1^{(g)})=(2,1)$. As $\mu/J$ decreases both slopes increase towards the conformal values $(q,2)=(4,2)$ of Eq.~\eqref{eq:towers_largeN}, while the intercepts stay at $p^{m,g}_{0}\approx1$. We compare the matter tower with the large-$N$ SD result~\cite{MQ,Lantagne2020,FloquetWormhole} and find qualitative agreement for $p_1$. Below $\mu\sim J$ the conformal gap scaling $E^{(m)}_{\rm gap}\sim\mu^{2/3}$ is recovered for every size studied, $N=14$--$20$, matching the SD result. The finite-$N$ ladders therefore follow the large-$N$ ones wherever the clusters remain resolved. The exact ratios of Eq.~\eqref{eq:towers_largeN} are only reached asymptotically as $\mu\to0$, and that regime lies below the coupling at which the clusters merge, $\mu\sim0.1$ for $N=16$. We return to the physical meaning of this merging in the Outlook.

\emph{ \color{blue}Coherent dynamics and revivals.--} This spectrum has direct dynamical consequences. Injecting a fermion into the ground state prepares the wavepacket $\ket{\Phi^{a}_{j}}=\sqrt2\,\chi^{a}_{j}\ket{E_0}$ which decomposes over the size sectors as $\tilde{P}_k\ket{\Phi^{a}_{j}}= c_{jk}^a\ket{\psi_{jk}^a}$; we suppress the $a,j$ labels from here on. Only odd $k$ contribute, with weights $w_k\approx|c_k|^{2}/2$ that peak at $k=1$ and decay along the envelope of Fig.~\ref{fig:towers}(a). Each size component obeys $\bra{\psi_k}H\ket{\psi_k} \simeq \bar{E}_k$ and $\sqrt{\bra{\psi_k}H^2\ket{\psi_k} - \bar{E}_k^2} \sim \Delta E_k$, so for $t\lesssim1/\Delta E_k$ it evolves as a single effective level, with phase $e^{-i\bar E_k t}$. Whenever $\Delta E_k$ is small, as for $k<k_c$ at large $N$, thermalization is evaded, realizing a subspace that exhibits weak ergodicity breaking~\cite{note_scar}\nocite{Serbyn2021,Moudgalya2022,Chakraborty2026}. This is in agreement with recent predictions that horizonless bulk geometries should generically produce holographic scars~\cite{Milekhin2024}.

The return amplitude of the injected fermion defines the transmission coefficient~\cite{Lantagne2020,Qi2020,Caceres2021}
\begin{equation}
T_{LL}(t)=\frac{1}{N}\Big|\sum_j\braket{\Phi^{L}_{j}}{\Phi^{L}_{j}(t)}
\Big|\simeq2\Big|\sum_{k\,\mathrm{odd}}w_k\,
e^{-i(\bar E_k-E_0)t}\Big|.
\label{eq:TLL}
\end{equation}
Because the centroids are nearly equispaced, the phases realign and $T_{LL}$ revives with period $t_{\rm rev}=2\pi/(p_1^{(m)}E^{(m)}_{\rm gap})$. The amplitude of these revivals is controlled by the distribution $w_k$. For $\mu\gg J$, most of the weight lives in the $k=1$ component and $T_{LL}$ flattens despite the towers being well resolved. The inter-boundary amplitude $T_{LR}(t)=\big|\sum_j\braket{\Phi_j^L}{\Phi_j^R(t)}\big|/N$ is delayed because $\bra{E_n^{\lambda_r}}\chi_j^R\ket{E_0}=i^{\,r}\bra{E_n^{\lambda_r}}\chi_j^L\ket{E_0}$. Since consecutive matter clusters alternate $r=1,3$, the factor $i^r$ amounts to a half-period time translation, $T_{LR}(t)\simeq T_{LL}(t-t_{\mathrm{rev}}/2)$, yielding the alternating revivals of Fig.~\ref{fig:towers}(d)~\cite{Caceres2021,Lantagne2020,Qi2020}. 

\emph{ \color{blue}Wormhole to black hole transition.--} At large $N$ a first-order transition separates the wormhole and two-black-hole saddles of the Euclidean path integral~\cite{MQ}. Its finite-$N$ signatures were identified through level statistics, thermodynamic observables, and overlap with the TFD state~\cite{GarciaGarcia2019}. Size clustering recasts this saddle competition as an interplay between operator-size energy and entropy. Organizing the partition function into size sectors, $Z(\beta)=\Tr e^{-\beta H}=\sum_k Z_k$ with $Z_k=\tilde N_k\,e^{-\beta\bar E_k}\big\langle e^{-\beta(E-\bar E_k)}\big\rangle_k$, defines a size-resolved free energy
\begin{equation}
\beta F_k(\beta)=-\log Z_{k}=-S_k+\beta \bar{E}_k-V_k(\beta),
\end{equation}
with $S_k=\log \tilde{N}_{k}$ the operator-size entropy, $\bar E_{k}$ the cluster centroid, and $V_k(\beta)=\log\left\langle e^{-\beta(E-\bar E_k)}\right\rangle_k$ the contribution from sector width. The partition function is dominated by the sector $k$ of smallest $F_k$, just as the large-$N$ path integral is dominated by the saddle of least action.

\begin{figure}[htbp]
    \centering
    
    \includegraphics[width=\linewidth]{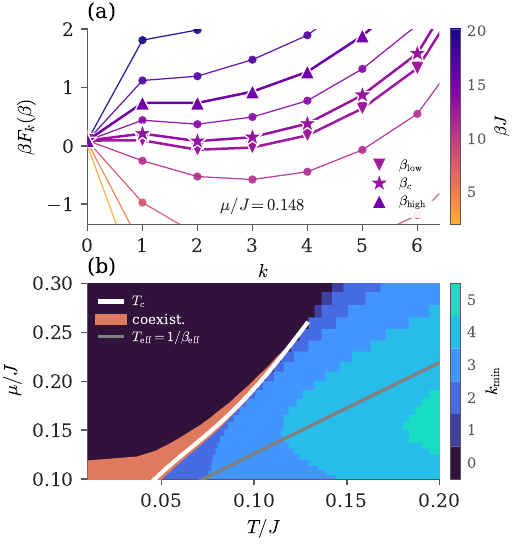}
    
    \caption{(a) $\beta F_k(\beta)$ for $N=16$, $\mu/J=0.148$, averaged over $50$ disorder realizations. For $0.13<\mu<0.27$ a double minimum structure develops between the marked $\beta_{\text{low}}$ and $\beta_{\text{high}}$, defining a coexistence region with the two minima degenerate at the critical $\beta_c$. For $\mu<0.13$ the second minimum survives at arbitrarily large $\beta$, a finite-$N$ effect of cluster mixing. (b) Resulting $(T,\mu)$ phase diagram, restricted to $\mu\ge0.10$ where the clusters stay resolved, colored by the size $k_{\text{min}}=\arg\min_k \beta F_k(\beta)$ at the free energy minimum.}
    \label{fig:transition_phase_diagram}
\end{figure}

At high temperature the entropy dominates, $\beta F_k\simeq -S_k$, which has a minimum at $k\sim N/2$. For small enough $\mu$ the clusters in that region of the spectrum have widths exceeding their spacing at any $N$, so the spectrum is indistinguishable from that of a generic chaotic system~\cite{GarciaGarcia2019}. We identify this as the finite-$N$ analogue of the black hole phase. On cooling, this minimum moves to smaller $k$, and eventually $\beta\bar E_k$ and $V_k(\beta)$ bend the profile of $\beta F_k$ enough to develop a competing minimum at $k=0$ [Fig.~\ref{fig:transition_phase_diagram}(a)]. The two become degenerate at the critical temperature $T_c=1/\beta_c$~\cite{note_BK}\nocite{Borgs1990}. Below $T_c$ the $k=0$ saddle dominates and the thermal state collapses onto the ground state, whose low-size excitations sustain the wormhole dynamics discussed above. Scanning $(T,\mu)$ yields the phase diagram of Fig.~\ref{fig:transition_phase_diagram}(b), a finite-$N$ microscopic precursor of the transition of Ref.~\cite{MQ}.

\emph{ \color{blue}Outlook.--} Operator-size clustering organizes the finite-$N$ spectrum; the conformal towers, the coherent revivals and the phase transition all follow from it. We close with two directions. First, we conjecture that the internal width $\Delta E_k$ is the finite-$N$ manifestation of gravitational backreaction. The $\mathcal{O}(1)$ energy injected by a single Majorana ceases to be negligible against the $\mathcal{O}(\mu N)$ binding energy, so the probe sources a positive-energy shockwave that stretches the wormhole and closes its throat~\cite{Gao2017,Maldacena2017}. This backreaction shows up as dephasing of the transmitted signal or, at smaller $\mu$, as the merging of the clusters and the collapse of the wormhole. Second, size resolution provides an operational diagnostic for quantum simulators. The resolved window $k<k_c$ and the coherence time $t^{k>1}_{\rm coh}\sim\sqrt N/k$ determine how many revivals a given device can sustain before dephasing sets in. The size distribution $W_k$ of the injected wavepacket, in turn, is directly measurable, and identifies which collective degrees of freedom a teleportation protocol may actually be probing.

\vspace{5pt}
\emph{ \color{blue}Acknowledgments.--} We thank Carolina Filgueira, Alexey Kitaev, Pavel Kos, Alexey Milekhin, Sebastián V. Romero, Thomas Schuster, and Alejandro Vilar for useful discussions.

This work has received financial support from the Xunta de Galicia (CIGUS Network of Research Centres and grant ED431C-2025/11), the European Union, the María de Maeztu grant CEX2023-001318-M funded by MICIU/AEI/10.13039/501100011033 and the Spanish Research State Agency AEI grant PID2023-152148NB-I00. The work of MB has been funded by Xunta de Galicia through the Programa de axudas á etapa predoutoral. The work of J.SS. was supported by FSE+ with the grant PRE2022-102163.

This research project was made possible through the access granted by the Galician Supercomputing Center (CESGA) to its supercomputing infrastructure. The supercomputer FinisTerrae III and its permanent data storage system have been funded by the NextGeneration EU 2021 Recovery, Transformation and Resilience Plan, ICT2021-006904, and also from the Pluriregional Operational Programme of Spain 2014-2020 of the European Regional Development Fund (ERDF), ICTS-2019-02-CESGA-3, and from the State Programme for the Promotion of Scientific and Technical Research of Excellence of the State Plan for Scientific and Technical Research and Innovation 2013-2016 State subprogramme for scientific and technical infrastructures and equipment of ERDF, CESG15-DE-3114. 

\bibliography{bib}

\onecolumngrid
\newpage


\setcounter{secnumdepth}{3}

\setcounter{equation}{0}
\setcounter{section}{0}
\setcounter{figure}{0}
\setcounter{table}{0}
\setcounter{page}{1}

\renewcommand{\theequation}{S\arabic{equation}}
\renewcommand{\thesection}{S\arabic{section}}
\renewcommand{\thefigure}{S\arabic{figure}}
\renewcommand{\thetable}{S\arabic{table}}

\renewcommand{\theHequation}{S\arabic{equation}}
\renewcommand{\theHsection}{S\arabic{section}}
\renewcommand{\theHfigure}{S\arabic{figure}}
\renewcommand{\theHtable}{S\arabic{table}}

{\centering
\large\bfseries
Supplemental Material: Size Operator and Spectral Clustering in the Two Coupled SYK Model
\par}

\section{Diagonalizing the Two Coupled SYK model \label{app:diagonalizing}}

Here we review the diagonalization of the Two Coupled SYK model, 
\begin{equation}
H(\mu) = H_{\text{SYK}}^L + H_{\text{SYK}}^R + \mu \Hint
\end{equation}
where
\begin{equation}
    H_{\text{SYK}}^a =i^{q/2}\sum_{j_{1}<\cdots<j_{q}}J_{j_{1}\cdots j_{q}}\chi_{j_{1}}^a\cdots\chi^a_{j_{q}}
\end{equation}
and
\begin{equation}
\Hint=i\sum_{j=1}^{N}\chi_{j}^{L}\chi_{j}^{R}.
\end{equation}

\subsection{Fock space and spectrum of \texorpdfstring{$\Hint$}{Hint}}
In order to define a Fock space, one has to define complex fermion operators by pairing Majorana modes. In this model, there are two reasonable options.

We can define a series of fermionic operators $f$ that couple even and odd Majorana modes,
\begin{equation}
    f^\dagger_j = \frac{1}{\sqrt{2}}\left(\chi_{2j-1} - i\chi_{2j}\right)\,, \quad f_j = \frac{1}{\sqrt{2}}\left(\chi_{2j-1} + i\chi_{2j}\right)\,, \quad j=1,\ldots, N \,.\label{eq:f_ops}
\end{equation}
Provided that $N$ is even, these operators do not mix left and right systems so they preserve the tensor product structure of the Hilbert space $\mathcal{H}=\mathcal{H}_L\otimes\mathcal{H}_R$, which is necessary for computing the left--right entanglement entropy $S_{\mathrm{LR}}$ in the main text.  

One could also choose to define a different set of operators that couple left and right Majorana modes,
\begin{equation}
    d_j = \frac{1}{\sqrt{2}}\left(\chi_{j}^L + i\chi_{j}^R\right) = \frac{1}{\sqrt{2}}\left(\chi_{j} + i\chi_{j+N}\right)\,, \quad j=1\ldots N\,. \label{eq:d_ops}
\end{equation}

The operators $d$ and $f$ are related as
\begin{equation}
    d_k = \frac{i^{(k+1)\bmod 2}}{2}\Big\{(-1)^{k+1}\big[f_{\left\lceil \frac{k}{2}\right\rceil}+ i f_{\left\lceil \frac{k}{2}\right\rceil+N/2}\big] + f_{\left\lceil \frac{k}{2}\right\rceil}^\dagger + i f_{\left\lceil \frac{k}{2}\right\rceil+N/2}^\dagger\Big\}, \qquad k=1,\dots,N.
\label{ftod}
\end{equation}

In the $d$ basis, $\Hint$ becomes a set of uncoupled harmonic oscillators,
\begin{equation}
    \Hint = i \sum_{j=1}^N \chi_j^L \chi_j^R = -\frac{N}{2} + \sum_{j=1}^Nd_j^\dagger d_j = -\frac{N}{2} + Q,
\end{equation}
where we have defined $Q\equiv \sum_{j=1}^Nd_j^\dagger d_j$. The energies of $\Hint$ are thus $E_k^{(0)}=k-N/2$ and its eigenstates $\ket{k,m}$ where $k\in \{0, \ldots ,N\}$ is the eigenvalue of $Q$, and $m\in\{1, \ldots, N_k\}$ accounts for the harmonic degeneracy.

The operator $Q$ has a direct physical meaning. Any operator $\mathcal{O}$ can be expanded in the complete basis of Majorana strings $\Gamma_a = \chi_{j_1} \chi_{j_2} \dots \chi_{j_{|a|}}$, where $a = \{j_1, j_2, \dots, j_{|a|}\}$, as $\mathcal{O} = \sum_a c_a \Gamma_a$. The size of the operator is mathematically defined as the weighted average string length $|a|$ in this expansion~\cite{QiStreicher2019}, 
\begin{equation}
    \text{Size}(\mathcal{O}) = \frac{\sum_a |a| |c_a|^2}{\sum_a |c_a|^2}.
\end{equation}
Under the vectorization map $\mathcal{O} \to \ket{\mathcal{O}}$ an operator acting on $N$ Majoranas is converted to a state in a doubled Hilbert space $\mathcal{H}_L \otimes \mathcal{H}_R$. The identity operator maps to the infinite-temperature thermofield double state. In the following, we will prove that this state is annihilated by the $d_j$ fermions and thus $Q$ measures the number of size excitations above it.

Consider the (single side) chirality $\gamma_5$ and charge $C$ operators
\begin{equation}
    \gamma_5 = 2^{N/2} i^{-N/2} \prod_{j=1}^N \chi_j\,, \quad C=\prod_{j=1}^{N/2} \chi_{2j-1}, \label{eq:gamma5C}
\end{equation}
then we define the Thermofield Double (TFD) state of a single SYK with eigenstates $\ket{e_n}$,
\begin{equation}
    \begin{aligned}
        \ket{\mathrm{TFD}(\beta)}&=Z_{\text{SYK}}^{-1/2}(\beta)2^{N/4}e^{-\beta(H_{\text{SYK}}^L+H_{\text{SYK}}^R)/4}\ket{\mathrm{TFD}(\beta=0)}\\
    &=Z_{\text{SYK}}^{-1/2}(\beta)\sum_n e^{-\beta e_n/2} \ket{e_n}_L\ket{\tilde{e}_n}_R\,, \label{eq:TFD}
    \end{aligned}
\end{equation}
where $\ket{\tilde{e}_n}_R=e^{-i\frac{\pi}{4}\gamma_5}C\ket{e_n}^*_R$ and $Z_{\text{SYK}}=\Tr (e^{-\beta H_{\text{SYK}}})$. The infinite temperature TFD is the ground state of $\Hint$~\cite{GarciaGarcia2019},
\begin{equation}
\begin{aligned}
    \ket{0, 0}=\ket{\mathrm{TFD}(\beta=0)} &= 2^{-N/4}\sum_n \ket{e_n}_L\ket{\tilde{e}_n}_R\\
     &=2^{-N/4}\sum_n\ket{e_n}_Le^{-i\frac{\pi}{4}\gamma_5}C\ket{e_n}^*_R.
    \label{eq:gs_hint}
\end{aligned}
\end{equation}
The Fock vacuum is the unique state annihilated by every $d_j$, so establishing Eq.~\eqref{eq:gs_hint} amounts to showing that $d_j\ket{\mathrm{TFD}(\beta=0)}=0$ for all $j$. Before proceeding with the proof, we must establish four key algebraic properties of our operator representation.

Let $\ket{\Omega} = \sum_n \ket{n} \otimes \ket{n}$ be the unnormalized maximally entangled state in a real computational basis (where $\ket{n}^* = \ket{n}$). For any matrix $A$ acting on the single system Hilbert space, the action of $A$ on the left subsystem is equivalent to the action of its transpose $A^T$ on the right subsystem. By expanding the action of $A$, we get
\begin{equation}
\begin{aligned}
    (A \otimes I)\ket{\Omega} &= \sum_n (A\ket{n}) \otimes \ket{n} = \sum_{n,m} \bra{m}A\ket{n} \ket{m} \otimes \ket{n} \\
    &= \sum_{m,n} A_{mn} \ket{m} \otimes \ket{n}\,.
\end{aligned}
\end{equation}
Similarly, acting on the right subsystem:
\begin{equation}
\begin{aligned}
    (I \otimes A^T)\ket{\Omega} &= \sum_m \ket{m} \otimes (A^T\ket{m}) = \sum_{m,n} \ket{m} \otimes \bra{n}A^T\ket{m} \ket{n} \\
    &= \sum_{m,n} A_{mn} \ket{m} \otimes \ket{n}\,.
\end{aligned}
\end{equation}
Therefore, we obtain the identity:
\begin{equation}
    (A \otimes I)\ket{\Omega} = (I \otimes A^T)\ket{\Omega}\,. \label{eq:max_entangled_id}
\end{equation}

Inverting Eq.~\eqref{eq:f_ops} yields the Majorana operators
\begin{equation}
    \chi_{2j-1} = \frac{1}{\sqrt{2}}(f_j + f_j^\dagger), \quad \chi_{2j} = \frac{1}{i\sqrt{2}}(f_j - f_j^\dagger)\,.
\end{equation}

Using the Pauli matrix representation $f_j = \left(\prod_{l<j} \sigma_z^{(l)}\right) \sigma_-^{(j)}$, we can write the Majoranas explicitly as:
\begin{equation}
    \chi_{2j-1} = \frac{1}{\sqrt{2}} \left(\prod_{l<j} \sigma_z^{(l)}\right) \sigma_x^{(j)}, \quad \chi_{2j} = \frac{1}{\sqrt{2}} \left(\prod_{l<j} \sigma_z^{(l)}\right) \sigma_y^{(j)}\,.
\end{equation}
Because $\sigma_x$ and $\sigma_z$ are real and symmetric matrices, their tensor products are also symmetric, meaning $\chi_{2j-1}^T = \chi_{2j-1}$. However, $\sigma_y$ is purely imaginary and antisymmetric ($\sigma_y^T = -\sigma_y$), which makes the entire even Majorana operator antisymmetric, $\chi_{2j}^T = -\chi_{2j}$. We can compactly express both cases as:
\begin{equation}
    \chi_j^T = (-1)^{j+1} \chi_j\,. \label{eq:majorana_transpose}
\end{equation}

The chirality operator $\gamma_5$ is proportional to the product of all $N$ Majorana operators. Consequently, $\gamma_5$ anticommutes with any single Majorana:
\begin{equation}
    \{\gamma_5, \chi_j\} = 0 \implies \gamma_5 \chi_j = -\chi_j \gamma_5\,.
\end{equation}
Thus,
\begin{equation}
\begin{aligned}
     e^{-i\frac{\pi}{4}\gamma_5} \chi_j &= 
    \left( \cos\left(\frac{\pi}{4}\right)I - i\sin\left(\frac{\pi}{4}\right)\gamma_5 \right) \chi_j \\
    &=\chi_j \left( \cos\left(\frac{\pi}{4}\right)I + i\sin\left(\frac{\pi}{4}\right)\gamma_5 \right) \\
    &= \chi_j e^{+i\frac{\pi}{4}\gamma_5}\,.
\end{aligned}
\end{equation}

The charge operator $C$ of Eq.~\eqref{eq:gamma5C} consists of exactly $N/2$ odd indexed Majorana operators. When commuting an arbitrary Majorana $\chi_j$ through $C$, we must consider two cases:
\begin{itemize}
    \item \textbf{If $j$ is odd ($j=2m-1$):} writing $C=\chi_1\chi_3\cdots\chi_{2m-1}\cdots\chi_{N-1}$, moving the trailing factor of $C\chi_{2m-1}$ leftward to meet its own copy in $C$ (past $N/2-m$ anticommuting factors) gives $C\chi_{2m-1}=(-1)^{N/2-m}[\chi_1\cdots\chi_{2m-3}\chi_{2m+1}\cdots\chi_{N-1}]$; the same manipulation on $\chi_{2m-1}C$, moving the leading factor rightward past the $m-1$ factors before it, gives $\chi_{2m-1}C=(-1)^{m-1}[\chi_1\cdots\chi_{2m-3}\chi_{2m+1}\cdots\chi_{N-1}]$. Comparing the two yields $C\chi_{2m-1}=(-1)^{N/2-1}\chi_{2m-1}C$.
    \item \textbf{If $j$ is even ($j=2m$):} $\chi_{2m}$ is not part of $C$. It anticommutes with all $N/2$ factors, producing a phase of $(-1)^{N/2}$, yielding $C \chi_{2m} = (-1)^{N/2} \chi_{2m} C$.
\end{itemize}

Both cases can be unified by considering the product $C \chi_j^T$. Using the transposition property derived in Eq.~\eqref{eq:majorana_transpose}:
\begin{itemize}
    \item For odd $j$: $C \chi_{2m-1}^T = C \chi_{2m-1} = (-1)^{N/2 - 1} \chi_{2m-1} C$\,.
    \item For even $j$: $C \chi_{2m}^T = - C \chi_{2m} = -(-1)^{N/2} \chi_{2m} C = (-1)^{N/2 - 1} \chi_{2m} C$\,.
\end{itemize}
In both scenarios, the sign from the transpose perfectly absorbs the difference in anticommutations, leading to the relation
\begin{equation}
    C \chi_j^T = (-1)^{N/2 - 1} \chi_j C \,.\label{eq:charge_commute}
\end{equation}

In the tensor product space $\mathcal{H}_L \otimes \mathcal{H}_R$, the $2N$ Majorana operators defined by the Jordan--Wigner transformation take the form:
\begin{equation}
    \chi_j^L=\chi_j \equiv \chi_j \otimes I, \qquad \chi_j^R=\chi_{j+N} \equiv \gamma_5 \otimes \chi_j
\end{equation}
The $\gamma_5$ operator acting on the left subspace arises naturally from the Jordan--Wigner string crossing the first $N$ sites, ensuring the global anticommutation relation $\{\chi_j, \chi_{k+N}\} = 0$. 

We can rewrite $\ket{\mathrm{TFD}(\beta=0)}$ using the maximally entangled state $\ket{\Omega}$ and the matrix $M = e^{-i\frac{\pi}{4}\gamma_5}C$, such that $\ket{\mathrm{TFD}(\beta=0)} \propto (I \otimes M) \ket{\Omega}$. Using Eq.~\eqref{eq:max_entangled_id}, we need to prove that 
\begin{equation}
    d_j\ket{\mathrm{TFD}(\beta=0)}=\frac{1}{\sqrt{2}}\left[ (I \otimes M \chi_j^T) + (i\gamma_5 \otimes I)(I \otimes \chi_j M) \right] \ket{\Omega} = \frac{1}{\sqrt{2}}\left[ I \otimes \left(M \chi_j^T + i \chi_j M \gamma_5^T \right) \right] \ket{\Omega} = 0\,.
    \label{app:annihilTFD}
\end{equation}
Eq.~\eqref{app:annihilTFD} requires the operator acting on the right subsystem to vanish,
\begin{equation}
    M \chi_j^T + i \chi_j M \gamma_5^T = 0 \,.\label{eq:annihilation_cond}
\end{equation}

To solve this, we apply the algebraic properties of the operators. We know that $\gamma_5^T = \gamma_5$ and $e^{-i\frac{\pi}{4}\gamma_5} \chi_j = \chi_j e^{+i\frac{\pi}{4}\gamma_5}$. The charge operator relates to chirality as $C \gamma_5 = (-1)^{N/2} \gamma_5 C$.

Substituting $M = e^{-i\frac{\pi}{4}\gamma_5}C$ into Eq.~\eqref{eq:annihilation_cond} and applying these commutation rules alongside Eq.~\eqref{eq:charge_commute}, we obtain
\begin{equation}
    (-1)^{N/2 - 1} \chi_j e^{+i\frac{\pi}{4}\gamma_5} C + i (-1)^{N/2} \chi_j e^{-i\frac{\pi}{4}\gamma_5} \gamma_5 C = 0\,.
\end{equation}
Dividing from the left by $(-1)^{N/2 - 1} \chi_j$ and from the right by $C$ (noting that $(-1)^{N/2}/(-1)^{N/2-1} = -1$) yields a condition purely in terms of the chirality matrix:
\begin{equation}
    e^{+i\frac{\pi}{4}\gamma_5} - i e^{-i\frac{\pi}{4}\gamma_5} \gamma_5 = 0\,.
\end{equation}
Using Euler's formula $e^{i\alpha\gamma_5} = \cos(\alpha)I + i\sin(\alpha)\gamma_5$ with $\alpha = \pi/4$, and knowing that $\gamma_5^2 = I$, the expression expands to:
\begin{equation}
\begin{aligned}
    \frac{1}{\sqrt{2}}(I + i\gamma_5) - i \left[ \frac{1}{\sqrt{2}}(I - i\gamma_5) \right] \gamma_5 &= \frac{1}{\sqrt{2}}(I + i\gamma_5) - \frac{1}{\sqrt{2}}(i\gamma_5 - i^2\gamma_5^2) \\
    &= \frac{1}{\sqrt{2}}(I + i\gamma_5) - \frac{1}{\sqrt{2}}(i\gamma_5 + I) = 0\,.
\end{aligned}
\end{equation}
This cancelation proves that $\ket{\mathrm{TFD}(\beta=0)}$ is annihilated by every $d_j$. Since the Fock vacuum is the unique such state, $\ket{\mathrm{TFD}(\beta=0)}=\ket{0,0}$ and is the ground state of $\Hint$.

\subsection{Symmetries}
Consider the operator
\begin{equation}
    \mathcal{U}(\theta) = e^{i\theta \Hint} = \prod_j\left(\cos\frac{\theta}{2}- 2\sin \frac{\theta}{2}\chi_j^L\chi_j^R\right) \quad \theta \in \left[0, 2\pi\right).
    \label{expU}
\end{equation}
Under the action of this operator the Majoranas transform as 
\begin{equation}
\begin{aligned}
& \mathcal{U}(\theta) \chi_j^L \mathcal{U}(\theta)^{\dagger}=\chi_j^L \cos \theta+\chi_j^R \sin \theta, \\
& \mathcal{U}(\theta) \chi_j^R \mathcal{U}(\theta)^{\dagger}=\chi_j^R \cos \theta-\chi_j^L \sin \theta .
\end{aligned}
\end{equation}

There are two special choices of $\theta$ that leave the total Hamiltonian $H(\mu)$ invariant for $q$ even. These are symmetries of the system.

For $\theta=\pi$, we have
\begin{equation}
\mathcal{U}(\pi) \chi_j^L \mathcal{U}(\pi)^{\dagger}=-\chi_j^L, \qquad \mathcal{U}(\pi) \chi_j^R \mathcal{U}(\pi)^{\dagger}=-\chi_j^R .
\end{equation}
This is equivalent to the chirality operator extended to the doubled Hilbert space $\mathcal{U}(\pi)\sim\gamma_5$.
On the other hand, for $\theta=\pi/2$,
\begin{equation}
\mathcal{U}(\pi/2) \chi_j^L \mathcal{U}(\pi/2)^{\dagger}=\chi_j^R, \qquad \mathcal{U}(\pi/2) \chi_j^R \mathcal{U}(\pi/2)^{\dagger}=-\chi_j^L\,,
\end{equation}
which leaves the Hamiltonian invariant by swapping left and right systems. Note that $\mathcal{U}(\pi/2)^2=\mathcal{U}(\pi)$.

We now analyze the spectrum of
\begin{equation}
\mathcal{U}(\pi/2) = e^{i\frac{\pi}{2}(-N/2 + Q)}= i^{-N/2} \, i^{Q}.
\end{equation}

It is diagonal in the Fock basis $\{\ket{k,m}\}$
\begin{equation}
    \mathcal{U}(\pi/2)\ket{k,m} = i^{-N/2}i^k \ket{k,m}.
\end{equation}
Since $i^k = i^{k \pmod 4}$, the eigenvalue depends exclusively on the equivalence class $r \equiv k \pmod 4$. Therefore, the distinct eigenvalues of $\mathcal{U}(\pi/2)$ can be compactly labeled by the index $r \in \{0, 1, 2, 3\}$ as
\begin{equation}
    \lambda_r = i^{-N/2}i^r.
\end{equation}

Thus, $\mathcal{U}(\pi/2)$ exhibits a significantly larger spectral degeneracy than the charge operator $Q$. While $Q$ resolves the Hilbert space into $N+1$ distinct eigenspaces $\mathcal{H}_k$, the symmetry operator groups these into only four sectors
$\mathcal{H}_r = \bigoplus_{k \equiv r \pmod 4} \mathcal{H}_k.$ Consequently, $\mathcal{U}(\pi/2)$ is diagonal in a far wider class of bases than $Q$. Any unitary transformation that mixes states across different particle number sectors $k$ but preserves the $\mathbb{Z}_4$ charge $r$ will break the diagonality of $Q$ while leaving $\mathcal{U}(\pi/2)$ diagonal. This greater basis flexibility allows the symmetry operator to remain diagonal even in representations where the total particle number is no longer a well-defined quantum number.

As such, the total Hilbert space can be decomposed in $4$ sectors as
\begin{equation}
\mathcal{H}=\bigoplus_{r=0}^{3}\mathcal{H}_r.
\end{equation}

The dimension of each subspace is given by
\begin{equation}
\begin{aligned} 
    \dim_r \equiv\dim \mathcal{H}_r &=\sum_{k \equiv r \, \pmod 4} \binom{N}{k} = \sum_{k=0}^N
\binom{N}{k} \delta_{k\equiv r\pmod4} = \frac{1}{4} \sum_{r^{\prime}=0}^3 i^{-rr^{\prime}} \sum_{k=0}^N \binom{N}{k}i^{r^{\prime}k} \\
        &= \frac{1}{4}\sum_{r^{\prime}=0}^3 i^{-rr^{\prime}}(1+i^{r^{\prime}})^N =\frac{1}{4}\left[2^N + i^{-r}(1+i)^N+i^{-3r}(1-i)^N \right]\\
        &= \frac{1}{4} \left[2^N + 2^{N/2}\left(e^{i\left(\frac{N\pi}{4} - \frac{r\pi}{2}\right)}+e^{-i\left(\frac{N\pi}{4} - \frac{r\pi}{2}\right)}\right)\right] \\
        &= 2^{N-2} + 2^{N/2-1} \cos{\left(\frac{N\pi}{4} - \frac{r\pi}{2}\right)}.
\end{aligned}
\end{equation}
Roughly, this gives that each of the subspaces is of dimension $\dim_r \sim 2^{N-2}$.

We can define the spectral projectors
\begin{equation}
P_r=\frac{1}{4}\sum_{r^{\prime}=0}^{3} \left[\lambda_r^{-1}\mathcal{U}(\pi/2)\right]^{r^{\prime}}.
\end{equation}
In the $d$-Fock basis, they become
\begin{equation}
    P_r = \sum_{k\equiv r\pmod4} \sum_{m=1}^{N_k} \ket{k,m} \bra{k,m}\,.
\end{equation}

The change of basis between the original fermions $f$ and the fermions $d$ can be written as
\begin{equation}
V_{\mathrm{int}}=P_{\mathrm{int}}\left(\bigotimes_{j=1}^{N/2}V_{\mathrm{local}}\right),
\end{equation}
where $P_{\mathrm{int}}$ is a fermionic permutation that reorders the basis as $\ket{n_1^L \cdots n_{N/2}^L n_1^R \cdots n_{N/2}^R} \to \ket{n_1^L n_1^R \cdots n_{N/2}^L n_{N/2}^R}$ and $V_{\mathrm{local}}$ is a $4\times4$ matrix that locally rotates each pair of fermions to a Bell basis.

Since the coupled SYK Hamiltonian $H(\mu)$ commutes with $\mathcal{U}(\pi/2)$, we can decompose $H(\mu) = \sum_{r=0}^3 P_r H(\mu) P_r \equiv \sum_{r=0}^3 H_r(\mu)$. Thus, the $\mathbb{Z}_4$ symmetry allows us to block-diagonalize $H(\mu)$ as
\begin{equation}
\begin{aligned}
    &H\ket{E_n^\lambda}=E_n^\lambda\ket{E_n^\lambda}\,, \\
    &\mathcal{U}(\pi/2)\ket{E_n^\lambda}=\lambda\ket{E_n^\lambda}\,.
\end{aligned}
\end{equation}
In many cases where the symmetry sector is not of particular relevance, we will drop the superscript $\lambda$. 

Using \eqref{eq:TFD}, \eqref{eq:gs_hint} and $[\mathcal{U}(\pi/2), H(\mu)]=0$,
\begin{equation}
\begin{aligned}
    \mathcal{U}(\pi/2)\ket{\mathrm{TFD}(\beta)}&=\mathcal{U}(\pi/2)Z_{\text{SYK}}^{-1/2}(\beta)e^{-\beta H(\mu=0)/4}\ket{0,0} \\
    &=Z_{\text{SYK}}^{-1/2}(\beta)e^{-\beta H(\mu=0)/4} \mathcal{U}(\pi/2) \ket{0,0} \\
    &= i^{-N/2}\ket{\mathrm{TFD}(\beta)}=\lambda_0\ket{\mathrm{TFD}(\beta)}.
\end{aligned}
\end{equation}
As such, TFD states belong to the symmetry sector $r=0$.

Because both $\ket{\mathrm{TFD}(\beta)}$ and $\ket{E_n}$ are eigenstates of $\mathcal{U}(\pi/2)$, $\braket{E_n^\lambda}{\mathrm{TFD}(\beta)} \propto \delta^{\lambda}_{\lambda_0}$ or $\ket{\mathrm{TFD}(\beta)}=\sum_n c_n\ket{E_n^{\lambda_{0}}}$. It is well established that the ground state of the coupled system closely approximates a Thermofield Double state at an effective temperature, $\ket{E_0(\mu)}\approx\ket{\mathrm{TFD}(\beff(\mu))}$~\cite{MQ,Cottrell2019,Alet2021, Caceres2021, GarciaGarcia2019,Schuster2025}. The ground state must belong to the same symmetry sector as the TFD state for any value of $\mu$, therefore $\lambda_{\text{GS}} =\lambda_{0}=i^{-N/2}$. 

This provides a microscopic justification for the success of the adiabatic cooling protocol recently proposed in Ref.~\cite{Schuster2025}. If $\mu(t)$ varies slowly, the time dependent Hamiltonian $H(\mu(t))$ continues to commute with $\mathcal{U}(\pi/2)$ at all times, restricting the evolution of the initial state $\ket{0,0}$ entirely to the $\lambda = i^{-N/2}$ subspace. For the adiabatic theorem to hold and successfully track the true ground state, the system must avoid level crossings and the relevant sector must remain gapped. The cluster analysis of the main text shows that the $k=0$ cluster stays separated from the $k=4$ cluster throughout the resolved window,so that the gap controlling the preparation stays open and non-adiabatic leakage remains suppresed.

\section{Scaling of the Internal Width of Low-\texorpdfstring{$k$}{k} Clusters}
\label{app:cluster_width}

Here we show analytically that low-size spectral clusters of the two coupled SYK model become increasingly sharp as $N$ grows. We first develop the argument in the perturbative regime $\mu \gg J$, where the calculation is fully controlled, and subsequently extend it to every $\mu$.

\subsection{Perturbative regime: \texorpdfstring{$\mu \gg J$}{mu >> J}}
To analyze the spectrum in the $\mu \gg J$ regime, we will apply degenerate perturbation theory on the operator
\begin{equation}
\tilde{H}(\mu)=\frac{H(\mu)}{\mu} = \Hint + \frac{J}{\mu}\left(\frac{H_{\text{SYK}}^L + H_{\text{SYK}}^R}{J}\right)=\Hint+\epsilon H_{2\text{SYK}},
\end{equation}
treating the SYK Hamiltonians $H_{2\text{SYK}} \equiv \frac{1}{J}\left(H_{\text{SYK}}^L+ H_{\text{SYK}}^R\right)$ as a perturbation that lifts the degeneracy of $\Hint$ with small coupling $\epsilon=J/\mu$.

As $\epsilon\to0$ ($\mu/J\to\infty$) the spectrum of $H(\mu)/\mu$ is dominated by $\Hint = Q-\frac{N}{2}$ and thus has energies $\tilde{E}_k^{(0)}=k-N/2$ and eigenstates $\ket{k,m}$. However, for any finite value of $\mu$ the $\epsilon H_{2\text{SYK}}$ term lifts the degeneracy and broadens each size sector into a cluster of eigenstates with width $\Delta \tilde{E}_k$.

Consider $q=4$, then
\begin{equation}
H_{2\text{SYK}} = -\frac{1}{J}\sum_{i<j<l<m} J_{ijlm} \left( \chi^L_i \chi^L_j \chi^L_l \chi^L_m + \chi^R_i \chi^R_j \chi^R_l \chi^R_m \right).
\end{equation}
By substituting the transformation to the $d$ fermions, the quartic products for the left and right sectors expand respectively as:
\begin{align}
\chi^L_i \chi^L_j \chi^L_l \chi^L_m &= \frac{1}{4} (d_i + d_i^\dagger)(d_j + d_j^\dagger)(d_l + d_l^\dagger)(d_m + d_m^\dagger), \\
\chi^R_i \chi^R_j \chi^R_l \chi^R_m &= \frac{1}{4} (d_i^\dagger - d_i)(d_j^\dagger - d_j)(d_l^\dagger - d_l)(d_m^\dagger - d_m).
\end{align}
Any individual monomial term resulting from these expansions can be characterized by the number of creation operators $p$ (where $0 \le p \le 4$) and annihilation operators $4-p$ it contains. 

In the expansion of the left sector, every operator combination enters with a positive sign. Conversely, due to the relative minus sign in $\chi^R_j = \frac{i}{\sqrt{2}}(d_j^\dagger - d_j)$, each annihilation operator $d$ introduces a factor of $-1$. Consequently, a monomial with $p$ creation operators carries a sign factor of $(-1)^{4-p} = (-1)^p$ in the right sector. 

When summing both terms to construct $H_{2\text{SYK}}$, the total amplitude for any specific monomial containing $p$ creation operators is modulated by the structural factor:
\begin{equation}
\chi^L_i \chi^L_j \chi^L_l \chi^L_m + \chi^R_i \chi^R_j \chi^R_l \chi^R_m \propto \left[ 1 + (-1)^p \right].
\end{equation}
This factor vanishes identically for odd values of $p \in \{1, 3\}$, which correspond to transitions in operator size of $\Delta k = \pm 2$. As such, only terms with $p \in \{0, 2, 4\}$ contribute, yielding that the $H_{2\text{SYK}}$ term can only induce transitions $\Delta k \in \{0, \pm 4\}$ as expected from the $\mathbb{Z}_4$ symmetry.

To quantify the internal broadening of each size sector systematically, we define the cluster width $\Delta \tilde{E}_k$ as the spectral radius of the perturbed subspace, expressing it as a perturbative expansion in powers of the parameter $\epsilon$:
\begin{equation}
  \Delta \tilde{E}_k = \epsilon \Delta \tilde{E}_k^{(1)} + \epsilon^2 \Delta \tilde{E}_k^{(2)} + \mathcal{O}(\epsilon^3),
  \label{eq:width_series}
\end{equation}
where $\Delta \tilde{E}_k^{(n)}$ denotes the contribution to the width arising at the $n$-th order of the expansion.

The leading order correction $\Delta \tilde{E}_k^{(1)}$ is determined by the spectral radius of the perturbation projected onto sector $k$, $H_k^{(1)} = P_k H_{2\text{SYK}} P_k$. This projection isolates the $\Delta k = 0$ ($p=2$) particle conserving channel. Expanding the Majoranas, 
\begin{equation}
    H_k^{(1)} = -\frac{1}{2J} \sum_{i<j<l<m} J_{ijlm} \sum_{\sigma} (-1)^{\text{sgn}(\sigma)} d_{\sigma(1)}^\dagger d_{\sigma(2)}^\dagger d_{\sigma(3)} d_{\sigma(4)},
\end{equation}
where $\sigma$ runs over the $\binom{4}{2} = 6$ ways of choosing which two of the indices $(i,j,l,m)$ carry the creation operators. The factor $(-1)^{\text{sgn}(\sigma)}$ is the signature of the corresponding permutation, and accounts for the fermionic signs picked up when reordering the operators.

Thus, $H_k^{(1)}$ is a sparse random matrix within the $N_k$-dimensional subspace with row connectivity $K(k)=\binom{k}{2}\binom{N-k}{2}$ and entries of typical size $v/2$, where $v \equiv \sqrt{\langle J^2\rangle}/J=\frac{\sqrt{6}}{N^{3/2}}$. The leading spectral radious of such a sparse block is set by its connectivity~\cite{Guhr1998, Rodgers1988, Kota2001}
\begin{equation}
\Delta \tilde{E}_k^{(1)} = \sqrt{K(k)} v = \sqrt{\binom{k}{2}\binom{N-k}{2}} \frac{\sqrt{6}}{N^{3/2}}.
\end{equation}
Notice that at $k=1$, the factor $\binom{1}{2} = 0$. Thus $\Delta \tilde{E}_1^{(1)} = 0$, meaning that the first cluster has no width at order $\epsilon$. In the small excitation limit ($1 < k \ll N$), the connectivity scales asymptotically as $\sqrt{K(k)} \simeq \frac{1}{2}kN$, reducing the first-order width to the scaling law,
\begin{equation}
\Delta \tilde{E}_k^{(1)} \sim \left( \frac{1}{2} k N \right) \frac{\sqrt{6} }{N^{3/2}} \sim \frac{k}{\sqrt{N}}. \label{eq:width_perturbative}
\end{equation}
In the main text and whenever we quote this result from here onwards, we will restrict it to the low excitation limit.

We now compute the second-order correction for the $k=1$ sector. It is convenient to introduce the off-diagonal blocks
\begin{equation}
W_{k\pm} = P_k\, H_{2\rm SYK}\, P_{k\pm4},
\end{equation}
which are the only channels allowed by the selection rules $\Delta k \in \{0,\pm4\}$. For $k=1$ the down channel is empty, $W_{1-}=0$, so the standard second-order effective Hamiltonian reduces to
\begin{equation}
H^{(2)}_{1} = \sum_{k'\neq 1}\frac{P_1 H_{2\rm SYK} P_{k'} H_{2\rm SYK} P_1}
{\tilde E^{(0)}_1-\tilde E^{(0)}_{k'}} = -\frac{1}{4}\, W_{1+}\, W_{1+}^{\dagger},
\label{eq:second_order}
\end{equation}
where we have used $\tilde{E}_1^{(0)} - \tilde{E}_5^{(0)} = -4$. The second-order block is therefore an exact Gram (Wishart-type) form rather than a Wigner matrix, and its spectrum is controlled by the statistics of the rectangular block $W_{1+}$, of dimensions $N\times N_5$. That structure is simple: the matrix element $(W_{1+})_{an}=\bra{1,a}H_{2\rm SYK}\ket{5,n}$ is nonzero only if the intermediate state $n$ contains the mode $a$, and it then reduces to a single bare coupling,
\begin{equation}
(W_{1+})_{an} \propto J_{ijlm}/J,
\qquad \{i,j,l,m\}=n\setminus\{a\}.
\end{equation}
Each row of $W_{1+}$ thus contains $\binom{N-1}{4}$ nonzero entries of typical size $w\equiv v/2$, with $w^2=3/(2N^3)$.

The Gram form separates the spectrum of $H^{(2)}_1$ into a shift and a spread, and the two are governed by different countings. The diagonal elements $(W_{1+}W_{1+}^{\dagger})_{aa}$ sum the squares of all $\binom{N-1}{4}$ couplings in a row, so their mean produces a uniform displacement of the cluster centroid of order $\binom{N-1}{4}w^2\sim N$, common to all modes to leading order. Only the part that depends on $a$ contributes to the width, and it is most easily read off the difference $(W_{1+}W_{1+}^\dagger)_{aa}-(W_{1+}W_{1+}^\dagger)_{bb}$: the $\binom{N-2}{4}$ couplings involving neither $a$ nor $b$ appear in both sums and cancel, leaving $\binom{N-2}{3}$ terms in each.

The off-diagonal elements instead run only over the virtual paths shared by $a$ and $b$, since the intermediate state must contain both modes, which again leaves $\binom{N-2}{3}$ choices. Hence
\begin{equation}
\big(W_{1+}W_{1+}^{\dagger}\big)_{ab}\propto \sum_{i<j<l} J_{aijl}\,J_{bijl},
\qquad
\sigma_{\rm off}=\sqrt{\binom{N-2}{3}}\;w^{2},
\qquad
\sigma_{\rm diag}=\sqrt{2}\;\sigma_{\rm off},
\label{eq:gram_stats}
\end{equation}
the extra $\sqrt2$ arising because the diagonal sums squares, $\mathrm{Var}(J^2)=2\langle J^2\rangle^2$, against $\mathrm{Var}(JJ')=\langle J^2\rangle^2$ off it. Both fluctuations are of the same order, $\sigma_{\rm diag},\sigma_{\rm off}\sim N^{-3/2}$, but a symmetric matrix of dimension $N$ has spectral radius of order $\sqrt N\,\sigma_{\rm off}\sim1/N$, so it is the off-diagonal entries that set the width. The $k=1$ cluster therefore broadens as $1/N$, in contrast with the $k/\sqrt N$ law of the higher clusters.

\begin{figure}[t]
  \centering
  \includegraphics[width=\linewidth]{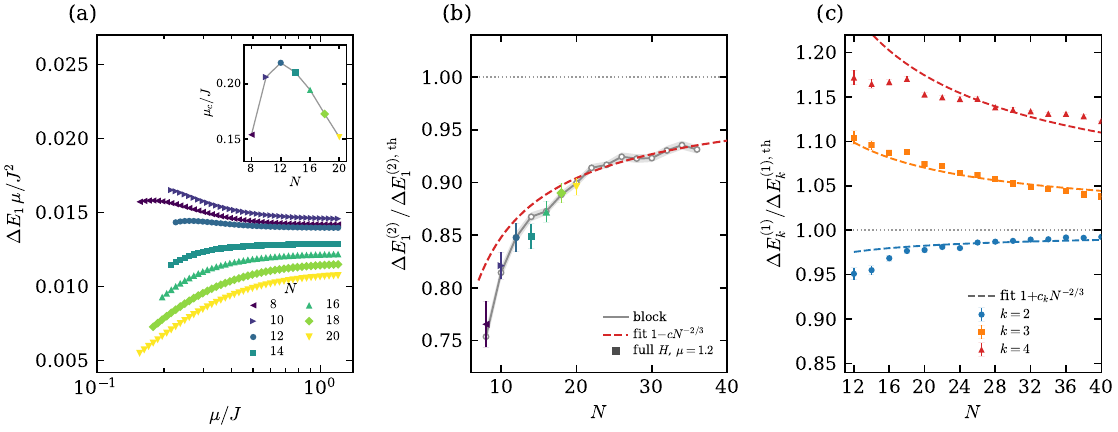}
  \caption{Internal width of the spectral clusters, computed as seed-averaged spectral radii of the cluster levels, for $q=4$. Solid markers of panels (a,b) are read off the spectrum of the full Hamiltonian $H(\mu)$ averaging over $50$ disorder realizations and reaching $N\in\{8,\ldots, 20\}$. Open markers in panel (b) and all data points of panel (c) are computed diagonalizing the projected operators, thus allowing us to reach $N=40$. \textbf{(a)} The $k=1$ width $\Delta E_1$ vs.\ $\mu$ follows a pure $\mu^{-1}$ law with
  \emph{no} plateau, confirming the analytical prediction $\Delta E_1^{(1)}=0$. Inset shows the value $\mu_c$ at which cluster $k=1$ mixes with the next cluster of the same symmetry sector $k=5$. For $N\geq12$, $\mu_c$ decreases; however, the system sizes accessible to exact diagonalization do not allow us to extract the exact $N$ dependence. \textbf{(b)} Second-order width rescaled by $\Delta E_1^{(2),\text{th}}=3J^2\sqrt{\binom{N-2}{3}}/(4 \mu N^{5/2})$. Numerical results agree with the analytical prediction up to finite-$N$ Tracy--Widom deviations decaying as $N^{-2/3}$. \textbf{(c)} First-order width rescaled by the analytical estimate, $\Delta E_k^{(1), \text{th}}=J\sqrt{6\binom{k}{2}\binom{N-k}{2}}/N^{3/2}$. Finite-$N$ corrections deviate the results from the analytical prediction; however, the $N^{-2/3}$ law does not seem to apply for the $N$ values studied.}
  \label{fig:appE_width}
\end{figure}

The $\mathcal{O}(1)$ edge constant follows as well, because Eq.~\eqref{eq:gram_stats} is precisely the variance profile of a GOE block. This is not spoiled by the entries being built from shared couplings: two off-diagonal elements sharing a row are uncorrelated, since $\langle(W_{1+}W_{1+}^{\dagger})_{ab}(W_{1+}W_{1+}^{\dagger})_{ac}\rangle$ would require pairing $J_{bijl}$ with $J_{cijl}$, which are independent for $b\neq c$. Once the common displacement of the diagonal is subtracted, the fluctuating part of the Gram block is thus a semicircle of radius $2\sqrt N\,\sigma_{\rm off}$. Including the factor $1/4$ of Eq.~\eqref{eq:second_order}, the width is $\tfrac12\sqrt N\,\sigma_{\rm off}=\tfrac12\sqrt N\,\sqrt{\binom{N-2}{3}}\,w^2$, and with $w^2=3/(2N^3)$ the prefactor is fixed with no free parameter,
\begin{equation}
\Delta\tilde E^{(2)}_1= A(N)\,\frac{\sqrt{\binom{N-2}{3}}}{N^{5/2}}\;\xrightarrow{\;N\to\infty\;}\;\frac{A_\infty}{\sqrt{6}\,N},
\qquad
A_\infty=\frac{3}{4}.
\label{eq:DE2}
\end{equation}
Direct diagonalization of Eq.~\eqref{eq:second_order} confirms this value and shows that it is approached from below, $A(N)=A_\infty\big(1-c\,N^{-2/3}\big)$ with $c\approx0.71$, the Tracy--Widom scaling of the extreme eigenvalue of a real symmetric block of dimension $N$. Restoring units,
\begin{equation}
\Delta E_1
= A(N)\,\frac{\sqrt{\binom{N-2}{3}}}{N^{5/2}}\,\frac{J^2}{\mu}
+\mathcal{O}\!\left(\frac{J^3}{\mu^2}\right) \;\xrightarrow{\;N\to\infty\;}\;\frac{3}{4\sqrt{6}N}\frac{J^2}{\mu} +\mathcal{O}\!\left(\frac{J^3}{\mu^2}\right)
\label{eq:DE1full}
\end{equation}
and
\begin{equation}
\Delta E_k = \mu\epsilon\Delta \tilde{E}_k^{(1)} +\mathcal{O}(\mu\epsilon^2)\sim \left( \frac{1}{2} k N \right) \frac{\sqrt{6} J}{N^{3/2}} +\mathcal{O}\left(\frac{J^2}{\mu}\right)\sim \frac{Jk}{\sqrt{N}}+\mathcal{O}\left(\frac{J^2}{\mu}\right), \qquad 1<k\ll N\label{eq:width_k>1}.
\end{equation}
These equations, tested against exact diagonalization in Fig.~\ref{fig:appE_width}, establish a parametric dichotomy in the leading order spectral broadening of the system. For generic multi-particle clusters ($k>1$), the dominant contribution arises at first order in perturbation theory, yielding a width that depends purely on the SYK coupling and the excitation size, with $1/\mu$ effects appearing only as subleading corrections. In contrast, since $\Delta E_1^{(1)}$ vanishes, its broadening is strictly a second-order effect parametrically suppressed by $\mu$. Consequently, in the strong coupling regime ($\mu \gg J$), the single particle excitation remains exceptionally narrow compared to the higher energy sectors. 

The Gram structure persists to all orders as the $k=1$ sector communicates with the rest of the spectrum only through the $k{=}1\!\leftrightarrow\!k{=}5$ channel, so every term of the series factorizes as $W_{1+}(\cdots)W_{1+}^{\dagger}$, with the inner factor dressing the internal dynamics of the upper band. The outer block fixes the same shift/spread separation as at second order, while the inner factor contributes $\mathcal{O}(1)$ in $N$, as each additional power of $v\sim N^{-3/2}$ is compensated by the combinatorial growth of the virtual band. This motivates the resummed form
\begin{equation}
\Delta E_1 = J\, f_1(N)\, g_1\!\left(\frac{J}{\mu}\right),
\qquad
f_1(N)=\frac{\sqrt{\binom{N-2}{3}}}{N^{5/2}}\sim\frac{1}{\sqrt6\,N},
\label{eq:resummed1}
\end{equation}
where $f_1(N)$ is the second-order scaling of Eq.~\eqref{eq:DE2}. We have verified this structure explicitly at second order. Its persistence at higher orders relies on the path counting of the inner factor remaining $\mathcal{O}(1)$. We will revisit this from the Feshbach--Fano viewpoint in the next subsection. For the higher clusters the first-order width dominates and the analogous form reads
\begin{equation}
\Delta E_k = \frac{Jk}{\sqrt N}\, g_k\!\left(\frac{J}{\mu}\right),\qquad k>1 .
\end{equation}

\subsection{Extension to the conformal regime}
\label{app:B2}

The perturbative result does not converge for $\mu \ll J$. To extend the estimate of $\Delta E_k$ beyond the perturbative limit, we apply the Feshbach--Fano partitioning formalism~\cite{Feshbach1964, Fano1961, Lowdin1962} to $H(\mu)$. Making use of the projectors $P_k$, and their orthogonal complement $P_k^\perp = 1 - P_k$, the eigenvalue problem can be recast as a self-consistent effective Hamiltonian,
\begin{equation}
H^{\rm eff}_{k}(E) = P_k H P_k + P_k H P_k^\perp \mathcal{R}(E) P_k^\perp H P_k,
\end{equation}
where $\mathcal{R}(E) = [E - P_k^\perp H P_k^\perp]^{-1}$ is the resolvent of the complementary subspace. This is just a rewriting so that every eigenvalue $E$ of $H(\mu)$ whose eigenvector $\ket{E}$ has support on sector $k$ solves $\det[E-H_k^{\rm eff}(E)]=0$. The price of this rewriting is that $H^{\rm eff}_k$ depends on $E$, so its eigenvalues $\lambda_j(E)$ are not eigenvalues of $H(\mu)$, which are instead the \emph{fixed points} $E_j=\lambda_j(E_j)$. Recovering a single level $E_j$ exactly would thus require diagonalizing its own matrix $H^{\rm eff}_k(E_j)$, a different one for each level of the cluster.

The diagonal block reduces to the same size preserving terms as in the perturbative expansion $P_k H P_k=\mu P_k\Hint P_k+JP_kH_{2\rm SYK }P_k=\mu E_k^{(0)}+JH_k^{(1)}$. Making use of the selection rules $\Delta k=0, \pm4$ one can write $P_kHP_k^{\perp}=JP_kH_{2\rm SYK }P_k^{\perp}=J(W_{k+}+W_{k-})$ so that
\begin{equation}
H^{\rm eff}_{k}(E) = \mu E_k^{(0)} +JH_k^{(1)}+ J^2 (W_{k+}+W_{k-})\mathcal{R}(E) (W_{k+}+W_{k-})^\dagger.
\end{equation}
The cross terms $W_{k\pm}\mathcal{R}(E)W_{k\mp}^\dagger$ vanish because, with hops of $\Delta k=\pm4$, sector $k$ is the only bridge between $k+4$ and $k-4$, and $\mathcal{R}(E)$ has no support on $k$. The effective Hamiltonian is therefore
\begin{equation}
H^{\rm eff}_k(E) \;=\; \mu E^{(0)}_k + J H^{(1)}_k + \Sigma_k(E),
\qquad
\Sigma_k(E) \;=\; J^2 \sum_{\pm} W_{k\pm}\, \mathcal{R}(E)\, W^{\dagger}_{k\pm}.
\label{eq:heff}
\end{equation}

Rather than solving the fixed point equations, we evaluate $H^{\rm eff}_k$ at the cluster centroid $\bar E_k$ and read the properties of cluster $k$ off the spectrum of $H^{\rm eff}_k(\bar E_k)$. We now show in which sense that spectrum is the spectrum of the cluster.

Away from its poles $\mathcal{R}(E)$ is Hermitian, so $\mathcal{R}(E)^2$ is positive semidefinite and
\begin{equation}
\partial_E \Sigma_k(E) \;=\; -\,J^2\sum_{\pm} W_{k\pm}\,\mathcal{R}(E)^2\,W^{\dagger}_{k\pm} \;\preceq\; 0 ,
\label{eq:dSigma}
\end{equation}
where $A\preceq0$ means $\bra{\psi}A\ket{\psi}\le0$ for every state. Since $\mu E^{(0)}_k+JH^{(1)}_k$ carries no energy dependence, the Hellmann--Feynman theorem transfers this sign to the $\lambda_j$ themselves, $\partial_E\lambda_j=\bra{j(E)}\partial_E\Sigma_k\ket{j(E)}\le0$, so they can only decrease as the energy grows.

The poles of $\mathcal{R}(E)$ are the eigenvalues of $P^\perp_k H P^\perp_k$, which are the levels of the neighboring sectors $k\pm4,k\pm8,\dots$ computed with sector $k$ removed. The energies of sector $k$ itself never appear, since $\mathcal{R}$ is built on the complement. Thus $\Sigma_k$ is a smooth function of $E$ throughout the cluster it describes, and $\bar E_k$ lies in the interior of a pole-free interval whose edges are the neighboring clusters. A cluster is resolved when it does not reach those edges.

Inside such an interval the eigenvalue conditions $f_j(E)\equiv\lambda_j(E)-E=0$ involve strictly decreasing functions, $\partial_E f_j\le-1$, so each $f_j$ vanishes at most once. The levels of $H(\mu)$ that make up the cluster are therefore in one-to-one correspondence with the $\lambda_j$ that cross the diagonal there, and the cluster carries the $N_k$ levels of the sector it grew out of.

The same slope controls how far those levels sit from the energy at which we evaluate the block. Let $\bar E_k$ and $E_j$ lie between the same two poles. Since $f_j(E_j)=0$, the mean value theorem gives $\lambda_j(\bar E_k)-\bar E_k=\partial_Ef_j(\xi)\,(\bar E_k-E_j)$ for some $\xi$ in between, and $|\partial_E f_j|\ge1$ turns this into a bound on the spectral radius $\Delta E_k=\max_j|E_j-\bar E_k|$ that we use as the width,
\begin{equation}
\begin{aligned}
\big|E_j-\bar E_k\big| &\;=\; \frac{\big|\lambda_j(\bar E_k)-\bar E_k\big|}{|\partial_Ef_j(\xi)|}\;\le\;\big|\lambda_j(\bar E_k)-\bar E_k\big| ,\\[2pt]
\Delta E_k &\;\le\; \max_j\big|\lambda_j(\bar E_k)-\bar E_k\big| .
\end{aligned}
\label{eq:width_bound}
\end{equation}
Evaluating at $\bar E_k$ can thus only overestimate how far each level lies from the centroid, and the spectral radius of $H^{\rm eff}_k(\bar E_k)$ is an upper bound on the width of cluster $k$.

The amount by which it overestimates is itself a physical quantity. Linearizing around $\bar E_k$ instead of bounding gives
\begin{equation}
\begin{aligned}
E_j-\bar E_k &\;=\; Z_j\,\big[\lambda_j(\bar E_k)-\bar E_k\big],\\[2pt]
Z_j &\;=\; \big[1-\partial_E\lambda_j(\bar E_k)\big]^{-1}\;\in\;(0,1] ,
\end{aligned}
\label{eq:Zfactor}
\end{equation}
and $Z_j$ is the weight that the exact eigenvector places on sector $k$. Indeed, reconstructing the full eigenvector from its projection, $\ket{E_j}\propto\big[1+\mathcal{R}(E_j)P^\perp_k H\big]P_k\ket{E_j}$, and using Eq.~\eqref{eq:dSigma},
\begin{equation}
\big\|P_k\ket{E_j}\big\|^{2} \;=\; \big[1-\partial_E\lambda_j\big]^{-1} \;=\; Z_j
\label{eq:Zweight}
\end{equation}
for normalized $\ket{E_j}$, with $\partial_E\lambda_j$ evaluated at the fixed point. Measured from the centroid, the spectrum of $H^{\rm eff}_k(\bar E_k)$ is therefore the spectrum of the cluster dilated level by level by $Z_j^{-1}$. Each level is placed too far from $\bar E_k$ by the inverse of the weight that its eigenstate has on sector $k$, so the estimate is exact when the eigenstates belong entirely to the sector and degrades when they do not. This is also when $k$ ceases to be a good label and the notion of a cluster of size $k$ loses its meaning. The hypothesis behind Eq.~\eqref{eq:width_bound} says the same thing in terms of energies: the absence of poles of $\mathcal{R}$ between the centroid and its levels means that no neighboring band reaches into the cluster.

Continuing $\mathcal{R}$ to $E\to E+i0^+$, the fluctuations of $H^{\rm eff}_k(\bar E_k)$ set the internal width $\Delta E_k$, while ${\rm Im}\,\Sigma_k$ measures the leakage of sector $k$ into its neighbors. When a neighboring band does reach the centroid, $\mathcal{R}$ acquires support at $E\simeq\bar E_k$, the weights $Z_j$ drop well below one and ${\rm Im}\,\Sigma_k$ becomes finite. The levels then stop forming a discrete block and the clusters cease to be resolved.

The width is governed by the off-diagonal elements of the block. For generic clusters ($k>1$), the dominant contribution is the first-order term $J H_k^{(1)}$, which generates a broadening $\Delta E_k^{(1)} \sim Jk/\sqrt{N}$ and carries no energy dependence at all, so the self-consistency discussed above only reaches the subleading part of the width. The off-diagonal elements of $\Sigma_k(E)$ provide those higher-order corrections. An off-diagonal element connects two different microstates through a common intermediate state, so the two must share their fermionic indices. This is the same counting that suppressed the off-diagonal entries in Eq.~\eqref{eq:DE2}, and it again leaves a subleading width of $\mathcal{O}(1/N)$. As such, we expect the general scaling of cluster width to be
\begin{equation}
 \Delta E_k \sim \frac{k}{\sqrt{N}} + \mathcal{O}(N^{-1}),
\end{equation}
where we have not included its dependence in $J$ and $\mu$. In the large-$N$ limit we therefore expect low-energy clusters to become degenerate so that $k$ remains a sharp label even though it is not conserved. Larger $k$ clusters do not become degenerate and instead mix with each other. As shown in the main text, we expect this dichotomy to underlie the wormhole (low $k$) and black hole (large $k$) phases of~\cite{MQ}.
The centroid energies defined as the mean energies of $H^{\rm eff}_k$ follow from the self-consistency equation
\begin{equation}
    \bar{E}_k \;\simeq\; \mu E^{(0)}_k + \overline{\Sigma}_k(\bar{E}_k),
\end{equation}
where we have taken into account that $H_k^{(1)}$ has zero mean. $\overline{\Sigma}_k$ is the disorder average of $\Sigma_k(E)$. We evaluate it in the self-consistent Born approximation (SCBA) keeping contractions in which the two couplings entering $\Sigma_k$ carry the same set of nodes and neglecting terms with correlations between the couplings in $W_{k\pm}$ and $\mathcal{R}(E)$ that contribute as $\mathcal{O}(1/N^2)$~\footnote{The approximation is the factorization $\langle W_{k\pm}\mathcal{R}(E)W_{k\pm}^\dagger\rangle\to
\langle W_{k\pm}W_{k\pm}^\dagger\rangle\,\langle \mathcal{R}(E)\rangle$: no disorder line connects the couplings in $W_{k\pm}$ to those inside $\mathcal{R}(E)$, i.e., non-crossing contractions with no vertex corrections. This is a Born / non-crossing self-energy, linear in a single dressed resolvent $\Gb$, and not the melonic $G^{q-1}$ self-energy of the standard SYK large-$N$ expansion; the two share a name only. The diagonality of $\Sigma_k$ is exact under disorder averaging since two different microstates $a \neq b$ of sector $k$ reach a common intermediate state $n$ through different couplings, so $(W_{k+})_{an}(W^*_{k+})_{bn}\propto\delta_{ab}$.}. The transition rate is controlled by the coupling variance $v^2 \sim N^{-3}$ and the combinatorial number of channels, yielding,
\begin{equation}
\overline{\Sigma}_k(E) \;=\; \frac{J^2 v^2}{4}
\left[\binom{N-k}{4}\,\Gb_{k+4}(E)
\;+\; \binom{k}{4}\,\Gb_{k-4}(E)\right],
\qquad
\Gb_{k'}(E) \equiv \frac{1}{N_{k'}}\,
\overline{\Tr\big[P_{k'} \mathcal{R}(E) P_{k'}\big]},
\label{eq:sigmabar}
\end{equation}
where $\Gb_{k'}(E)$ is the disorder-averaged local resolvent of the size sector $k'$. The binomials simply count the available transition channels, and all the strong-coupling physics resides in
$\Gb_{k\pm4}(E)$, which we do not expand. Note that $\partial_E\overline{\Sigma}_k$ is then proportional to the identity on the sector, so the weights $Z_j$ of Eq.~\eqref{eq:Zfactor} are common to the whole cluster and the energy dependence rescales it as a block, without distorting its shape or its internal level statistics. For $k \ll N$ the up-channel can be expanded as,
\begin{equation}
\frac{J^2 v^2}{4}\binom{N-k}{4} = \frac{J^2 N}{16} -  \frac{3J^2}{8}-\frac{J^2 k}{4}
+ \mathcal{O}\!\big(J^2 k^2/N\big) \sim \mathcal{O}(N) -\mathcal{O}(1)- \mathcal{O}(1) k + \mathcal{O}(N^{-1}) k^2,
\end{equation}
while the down-channel coefficient is suppressed by $(k/N)^3$ relative
to the linear term. This expansion naturally separates the diagrammatic contributions by their large-$N$ scaling. The $\mathcal{O}(N)$ piece corresponds to disconnected vacuum bubbles and, together with the $-3J^2/8$ term, produces a global energy shift common to all low-$k$ sectors that cancels when computing cluster spacings. The $\mathcal{O}(1)k$ contribution represents the leading connected contribution that survives the thermodynamic limit. It drives the asymmetric repulsion from the upper bands and renormalizes the harmonic spacing into the emergent gap $\varepsilon \sim J(\mu/J)^{2/3}$ (for $q=4$). Finally, the $\mathcal{O}(N^{-1})$ and higher-order terms encode subleading corrections arising from finite-size orbital overlaps. Thus from the spacing of clusters of the same symmetry sector $\Delta_k = \bar E_{k+4} - \bar E_k$, we get
\begin{equation}
\frac{\Delta_k}{4} = \mu - \frac{J^2}{4}\Gb_{k+4}\left(\bar E_k\right).
\label{eq:spacing}
\end{equation}
Two assumptions are used here. First, the $k$-independent pieces cancel between the two centroids only if consecutive rungs see the same local resolvent, $\Gb_{k+8}(\bar E_{k+4})\simeq\Gb_{k+4}(\bar E_k)$, that is if the ladder is locally uniform over the low-$k$ window. This holds while the clusters are resolved and evenly spaced, and it is the only place where a $k$ dependence of $\Gb$ is discarded. Second, the down channel is dropped, since it is suppressed by $(k/N)^3$.

This interpolates between the $\mu \gg J$ and $\mu \ll J$ regimes, but it fixes the spacing only once $\Gb_{k+4}$ is known, and $\Gb_{k+4}$ obeys its own equation involving $\Gb_k$ and $\Gb_{k\pm8}$: the sectors form a coupled chain in size space, which we do not solve in general.

In the perturbative limit, $\mu\gg J$, the adjacent band is narrow and the chain closes on its bare form, $\Gb_{k+4}(\bar E_k)\to-1/\Delta_k\approx-1/(4\mu)$. In this regime, the cluster spacing recovers the harmonic behavior with a small perturbative correction, yielding $\Delta_k/4\approx\mu+J^2/(16\mu)$.

As the coupling is lowered into the holographic window, $\mu \ll J$, the SYK interactions alter the low-frequency behavior of the resolvent. The spacing is then no longer set by the bare slope of the size potential, but by the local spectral density of the size band encoded in $\Gb_{k+4}$. In the large-$N$ limit this is known~\cite{MQ, Lantagne2020, Caceres2021} to follow $\varepsilon \sim \mu^{2/3}$ for $q=4$, in agreement with the results shown in the main text.

At finite $N$ and small enough $\mu$, while the conformal gap attempts to shrink according to the $\mu^{2/3}$ power law, the internal width of the individual clusters, $\Delta E_k \sim Jk/\sqrt{N}$, does not compress at the same rate. Progressively, starting from the middle of the spectrum, adjacent clusters overlap and hybridize, the centroid falls within the continuous support of the neighboring band, and the resolvent $\Gb_{k+4}$ acquires a finite imaginary part. The centroids cease to be well-defined isolated poles and dissolve into a continuous profile characteristic of RMT. The calculations done here are not valid in that regime since the SCBA does not hold. 

For any finite-$N$, the fractional scaling $\mu^{2/3}$ has a strict analytical cutoff before reaching $\mu \to 0$. A pure asymptotic scaling of the form $\bar{E}_k \sim \mu^{2/3}$ would imply a diverging derivative at the origin, $d\bar{E}_k/d\mu \sim \mu^{-1/3} \to \infty$. Yet, via the Hellmann--Feynman theorem, the energy derivative of eigenstates is strictly bounded in finite systems. Since $H_{\rm int} = Q - N/2$, and the eigenvalues of the size operator are bounded so that $0 \le \langle Q \rangle \le N$, the derivative of the energy with respect to the coupling is
\begin{equation}
\frac{dE_n}{d\mu} = \bra{E_n} \Hint \ket{ E_n} = -\frac{N}{2} + \bra{E_n} Q \ket{E_n} \le \frac{N}{2}.
\end{equation}
This implies that the scaling cannot be sustained for $\mu\lesssim (2/N)^3=\mathcal{O}(N^{-3})$. The same reasoning can be applied to the gap
\begin{equation}
\frac{d\Delta_k}{d\mu}=\frac{d}{d\mu}\big(\bar E_{k+4}-\bar E_k\big)=\langle Q\rangle_{k+4}-\langle Q\rangle_{k},
\end{equation}
whose right-hand side is bounded by $N$ as well. The conformal spacing $\Delta_k\sim\mu^{2/3}$ must therefore saturate below the same scale $\mu\sim\mathcal{O}(N^{-3})$. In practice the clusters merge long before that, so this is never the limiting effect.

\section{Dressed size basis and soft clustering}\label{app:dressing_clustering}

As derived in the previous section, the SYK interactions broaden the degenerate energy levels of the harmonic limit into clusters with an internal width $\Delta E_k \sim k/\sqrt{N}$. Because this width is suppressed by the system size, $k$ remains a well-behaved, approximately conserved quantum number at the edges of the spectrum even away from the $\mu \to \infty$ limit. Here we construct the operators that make the label $k$ precise at finite $\mu$. We construct first the exact ones, then the tractable approximation used throughout the paper.

At finite $\mu$ the ground state $\ket{E_0(\mu)}$ is no longer the Fock state annihilated by the $d_j$, so counting size above it requires adapted modes. The exact ones are obtained by transporting the bare $d_j$ along the adiabatic flow generated by
\begin{equation}
    U(\mu, \mu_0) = \mathcal{T}\exp\left(-i\int^{\mu}_{\mu_0}\!\mathcal{A}(\mu')\,d\mu'\right),
    \qquad
    [H(\mu),i\mathcal{A}(\mu)] = \partial_\mu H(\mu)=\Hint,
    \label{eq:adiabatic_flow}
\end{equation}
where $\mathcal{A}(\mu)$ is the adiabatic gauge potential. $U(\mu,\mu_0)$ maps eigenstates of $H(\mu_0)$ onto eigenstates of $H(\mu)$. Taking $\mu_0=\infty$, the size operator $Q$ and the projectors to its eigenspaces $P_k$ can be extended to finite $\mu$,
\begin{equation}
    \mathcal{Q}(\mu) = U Q U^\dagger, \qquad \mathcal{P}_k(\mu) = U P_k U^\dagger.
    \label{eq:transported}
\end{equation}
Because $U$ is unitary, $U d_j U^\dagger$ obeys the canonical anticommutation relations, its adjoint is $Ud_j^\dagger U^\dagger$, and the $\mathcal{P}_k(\mu)$ are mutually orthogonal Hermitian projectors resolving the identity. 

This construction is exact and impractical. Since the SYK terms are $q$-body, each commutator with $\mathcal{A}(\mu)$ in the expansion of the exponential raises the fermionic string by $q-2$, so the operator basis needed to represent $Ud_jU^\dagger$ grows exponentially in $N$ and neither a closed form nor a numerical integration is feasible beyond the smallest sizes.

We therefore approximate the flow, keeping its structure but replacing its generator. The true ground state is very close to a thermofield double, $\ket{E_0(\mu)}\approx\ket{\mathrm{TFD}(\beff(\mu))}$~\cite{MQ,Cottrell2019,Alet2021,Caceres2021,GarciaGarcia2019,Schuster2025,Khor2026}, with the effective temperature fixed by the best overlap,
\begin{equation}
    \beff(\mu) = \arg\max_\beta |\braket{ E_0(\mu)}{\mathrm{TFD}(\beta) }|^2.
    \label{eq:beff}
\end{equation}
The TFD definition of Eq.~\eqref{eq:TFD} motivates replacing $\mathcal{A}(\mu)\to -i\partial_\mu\beff(\mu)(H^L_{\text{SYK}}+H^R_{\text{SYK}})/4$. Since $H^L_{\text{SYK}}$ and $H^R_{\text{SYK}}$ commute, the path ordering drops out and, using $\beff(\mu_0\to\infty)=0$, the unitary $U(\mu,\mu_0)$ is replaced by the non-unitary similarity transformation $S(\mu)=e^{-\beff(\mu)(H^L_{\text{SYK}}+H^R_{\text{SYK}})/4}$. This defines the dressed fermionic operators
\begin{equation}
    d_j(\mu) = S d_j S^{-1}, \quad \tilde{d}_j(\mu) = S d_j^\dagger S^{-1}, \qquad \tilde{d}_j(\mu)\neq d_j^\dagger(\mu).
    \label{eq:dressed_ops}
\end{equation}
Equivalently, they are Majoranas evolved to imaginary time $\tau=-\beff/4$ under the Hamiltonian $H_{\text{SYK}}^L+H_{\text{SYK}}^R$,
\begin{equation}
    d_j(\mu) = \frac{1}{\sqrt{2}}\left[\chi_j^L(\tau)+i\,\chi_j^R(\tau)\right].
    \label{eq:imag_time}
\end{equation}
Using Eq.~\eqref{app:annihilTFD}, the TFD is the dressed $k=0$ vacuum as $d_j(\mu)\ket{\mathrm{TFD}(\beff)}=0$. Acting with the $\tilde d_j(\mu)$ then builds the dressed states
\begin{equation}
    \ket{k,m}_{\beff} = S\ket{k,m}=Z_{\mathrm{SYK}}^{1/2}(\beff)\,2^{-N/4}\;\tilde{d}_{j_1}(\mu)\cdots\tilde{d}_{j_k}(\mu)\ket{\mathrm{TFD}(\beff)} .
    \label{eq:dressed_states}
\end{equation}
Because $S$ is not unitary, these dressed states are neither normalized nor mutually orthogonal,
\begin{equation}
    {}_{\beff}\braket{k,m}{k',m'}_{\beff} \propto \bra{k,m} e^{-\beff (H_{\text{SYK}}^L+H_{\text{SYK}}^R)/2}\ket{k',m'} \neq \delta_{kk'}\delta_{mm'}.
    \label{eq:gram}
\end{equation}
The dressed number operator is
\begin{equation}
    \tilde{Q}(\mu) = S Q S^{-1} = \sum_{j=1}^N \tilde{d}_j(\mu)\,d_j(\mu),
    \label{eq:Q_dressed}
\end{equation}
which is non-Hermitian but isospectral with $Q$ (its eigenvalues are still the integers $k$). It approximately counts dressed size above $\ket{E_0}$.

\begin{figure}
    \centering
    \includegraphics[width=\linewidth]{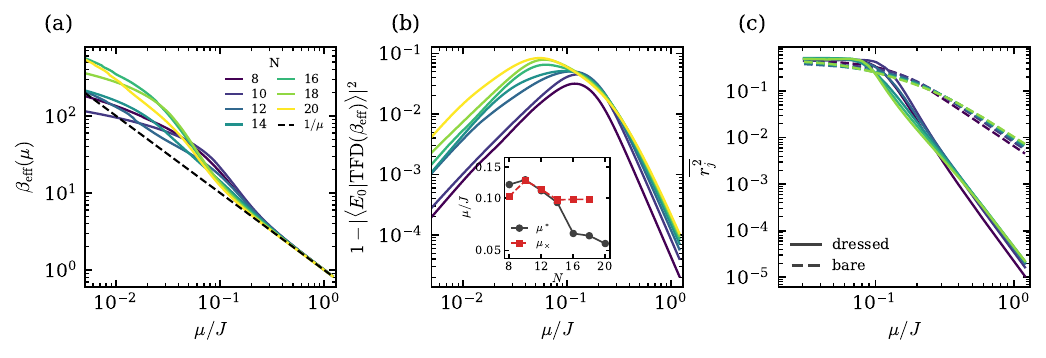}
    \caption{Accuracy of the thermal dressed vacuum. (a) Effective inverse temperature $\beff(\mu)$ of Eq.~\eqref{eq:beff} for $N=8$--$20$, compared with the large-$\mu$ estimate $\beff\approx1/\mu$ (dashed). (b) TFD-overlap deficit $1-|\braket{E_0}{\mathrm{TFD}(\beff)}|^2$ versus $\mu$. The deficit stays small throughout the clustered regime. It is non-monotonic and peaks at the few-percent level, at a coupling we call $\mu^*(N)$. The inset plots two scales against $N$: $\mu^*$, the position of this deficit peak, and $\mu_\times$, the coupling at which the dressed and bare residuals of panel (c) cross. Both scales fall in the window $\mu\approx0.06$--$0.13$ where the clusters merge. (c) Annihilation residual $\overline{r_j^2}$, averaged over the modes $j$ and over $50$ disorder realizations, shown for $N=8$--$20$. We plot it for the dressed operators $d_j(\mu)$ and for the bare $d_j$. The dressed residual is $r_j^2=\lVert d_j(\mu)\ket{E_0}\rVert^2/(\lVert d_j(\mu)\ket{E_0}\rVert^2+\lVert \tilde d_j(\mu)\ket{E_0}\rVert^2)$, and for the bare operators it reduces to the occupation $\bra{E_0}d_j^\dagger d_j\ket{E_0}$. Throughout the resolved regime the dressed residual lies orders of magnitude below the bare one. Below $\mu_\times$ the two curves cross and the dressing stops helping. This crossing marks the operator-level breakdown of the single-temperature TFD description of the ground state.}
    \label{fig:appC_vacuum}
\end{figure}

Panel~(a) of Fig.~\ref{fig:appC_vacuum} shows $\beff(\mu)$ while panel~(b) measures the \emph{global} quality of the vacuum through the TFD-overlap deficit $1-|\braket{E_0}{\mathrm{TFD}(\beff)}|^2$. It stays small throughout the clustered regime but is non-monotonic, peaking at a coupling $\mu^*(N)$ that drifts from $\mu^*\approx0.12$ at $N=8$ to $\mu^*\approx0.055$ at $N=20$, with a peak height that grows slowly with $N$. This growth is expected rather than alarming: the overlap is a global fidelity, so a small error per degree of freedom accumulates over an extensive number of modes into an $\mathcal{O}(1)$ deficit, as is generic for coupled parent Hamiltonians of this type~\cite{Khor2026}. The global overlap is therefore not a clean figure of merit at large $N$, which motivates an \emph{intensive}, operator-level probe. Panel~(c) provides it through the annihilation residual $\overline{r_j^2}$, with $r_j^2=\lVert d_j(\mu)\ket{E_0}\rVert^2/(\lVert d_j(\mu)\ket{E_0}\rVert^2 +\lVert \tilde d_j(\mu)\ket{E_0}\rVert^2)$, a per-mode quantity that vanishes for a perfect vacuum and reduces, for the bare operators, to the occupation $\bra{E_0}d_j^\dagger d_j\ket{E_0}$. In the resolved regime the dressing works as intended: the dressed residual lies orders of magnitude below the bare one ($2\times10^{-5}$ versus $6\times10^{-3}$ at $\mu=1.2$, and $3\times10^{-3}$ versus $7\times10^{-2}$ at $\mu\approx0.3$, essentially independent of $N$), so $d_j(\mu)$ annihilates $\ket{E_0}$ to high accuracy mode by mode. Below a crossover coupling $\mu_\times(N)$ the dressed residual overtakes the bare one and the dressing ceases to help. The inset of panel~(b) shows that $\mu_\times$ tracks the deficit peak $\mu^*$ closely up to $N=14$, beyond which it becomes nearly $N$-independent: $\mu_\times=0.098(1)$ for $N=14$--$18$ and $0.092(1)$ at $N=20$, while $\mu^*$ keeps decreasing by roughly a factor of two over the same range. Thus, both diagnostics locate the same breakdown, in the narrow window $\mu\approx0.06$--$0.13$ (Fig.~\ref{fig:appC_vacuum}). Throughout the resolved window $\mu>\mu_c(N)$ where the clustering analysis is performed, the thermal vacuum is thus accurate at the level of local, intensive observables.

The non-monotonicity of the deficit is an artifact of the TFD family itself. As $\beta\to\infty$ the thermofield double collapses to the product $\ket{\mathrm{TFD}(\beta\to\infty)}=\ket{e_0}_L\ket{\tilde e_0}_R$, which is precisely the $\mu\to0$ limit of $\ket{E_0(\mu)}$. Below $\mu^*$ the recovering overlap therefore certifies nothing: it records that the two states being compared have collapsed onto the same decoupled product. Note that a growing $\beff$ does not by itself invalidate the dressed description---$d_j(\mu)$ annihilates $\ket{\mathrm{TFD}(\beff)}$ by construction at any $\beff$. What degrades is the dressing map as $\beff(\mu)$ diverges: $S$ becomes singular and the oblique projectors ill-conditioned. This is a loss of the diagnostic, not necessarily of the clusters themselves, whose merging occurs when the inter-cluster spacing falls below the width $\Delta E_k\sim Jk/\sqrt{N}$. That both breakdowns occur in the same window, $\mu\approx0.06$--$0.13$ at the sizes studied, is why we restrict the
quantitative analysis to $\mu>\mu_c(N)$.

\begin{figure}
    \centering
    \includegraphics[width=\linewidth]{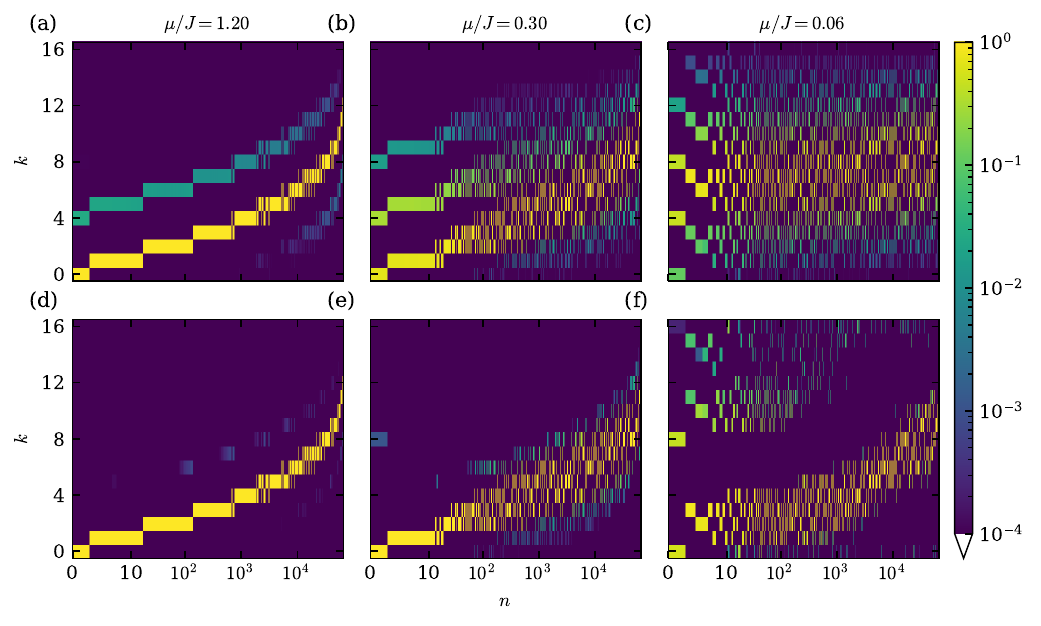}
    \caption{Bare and dressed cluster weights for $N=16$, $q=4$ (one disorder realization), at the three couplings $\mu=1.20,\,0.30,\,0.06$ of the main text. Panels (a)--(c): bare weights $W^b_k(n)=\bra{E_n}P_k\ket{E_n}$. Panels (d)--(f): dressed weights $W_k(n)$. The logarithmic scale on the $x$ axis expands the low-energy region. The bare weights spread eigenstates over several sectors (at $\mu=0.06$ only $11\%$ of $\ket{E_0}$ remains at $k=0$); the dressed weights keep them better resolved.}
    \label{fig:appC_bare_vs_dressed}
\end{figure}

The effect of the dressing on the cluster membership is shown directly in Fig.~\ref{fig:appC_bare_vs_dressed}, which compares the weights $W_k(n)$ computed with the bare projectors and with the dressed ones. The bare projectors smear the ground state and the low-lying states over many $k$, whereas the dressed projectors concentrate $\ket{E_0}$ at $k=0$ and keep the clusters sharp.

\subsection{Spectral clustering analysis}
This subsection details the weighting scheme used to assign eigenstates to clusters.

To resolve the dressed size sectors we use the oblique projectors
\begin{equation}
    \tilde{P}_k(\mu) = S P_k S^{-1}, \qquad P_k=\sum_{m=1}^{N_k}\ket{k,m}\bra{k,m},
    \label{eq:oblique_proj}
\end{equation}
and define the fractional membership weight of an exact eigenstate $\ket{E_n}$ to cluster $k$ as
\begin{equation}
    \mathcal{W}_k(n) = \bra{E_n}\tilde{P}_k(\mu)\ket{E_n} = \delta_{\lambda,\lambda_k}\,\bra{E^\lambda_n}\tilde{P}_k(\mu)\ket{E^\lambda_n}.
    \label{eq:W_def}
\end{equation}
The second equality uses that $\tilde P_k(\mu)$ has support only on the $\mathbb{Z}_4$ sector with $\lambda_k=i^{k-N/2}$, so an eigenstate contributes to cluster $k$ only if it lives in the matching charge channel. Substituting Eq.~\eqref{eq:oblique_proj}, Eq.~\eqref{eq:W_def} factorizes into a right vector, which is just the dressed state of Eq.~\eqref{eq:dressed_states}, $\ket{R_{k,m}}=\ket{k,m}_{\beff}=S\ket{k,m}$, and a left vector $\bra{L_{k,m}}=\bra{k,m}S^{-1}$, its biorthogonal dual so that $\braket{L_{k,m}}{R_{k,m'}}=\delta_{mm'}$,
\begin{equation}
    \mathcal{W}_k(n) = \sum_m \braket{E_n}{R_{k,m}}\braket{L_{k,m}}{E_n}=\sum_{m,m^\prime} \braket{E_n}{R_{k,m}}\braket{R_{k,m}}{L_{k,m^\prime}}\braket{L_{k,m^\prime}}{E_n}.
    \label{eq:KD}
\end{equation}
Because $S$ is Hermitian but not unitary, $\braket{L_{k,m}}{E_n}\neq\braket{E_n}{R_{k,m}}^{*}$, so the terms of Eq.~\eqref{eq:KD} can be complex. This is precisely the structure of a Kirkwood--Dirac quasiprobability~\cite{Hofmann2012,YungerHalpern2018} for a degenerate subspace, where the standard projector is replaced by the biorthogonal resolution $\sum_m \ket{R_{k,m}}\bra{L_{k,m}}$.

The model has an antiunitary symmetry $T=CK$, with $C$ given in Eq.~\eqref{eq:gamma5C}, $K$ complex conjugation and $T^2=+1$~\cite{GarciaGarcia2019}. Since $T^2=+1$, there is a basis of $T$-invariant states in which $T$ acts as pure complex conjugation. There the two SYK Hamiltonians and $\Hint$ are real symmetric matrices; hence so are $S$ and the spectral projectors $P_k$ of the real symmetric operator $\Hint$. The weights $\mathcal{W}_k(n)$ are therefore real. They obey the sum rules
\begin{equation}
    \sum_{k=0}^N \mathcal{W}_k(n) = 1, \qquad \sum_n \mathcal{W}_k(n) = \Tr\tilde P_k = \binom{N}{k}=N_k,
    \label{eq:sum_rules}
\end{equation}
the first by completeness of the $\tilde P_k$ and the second because the dressed sector reproduces the harmonic degeneracy.

\begin{figure}
    \centering
    \includegraphics[width=\linewidth]{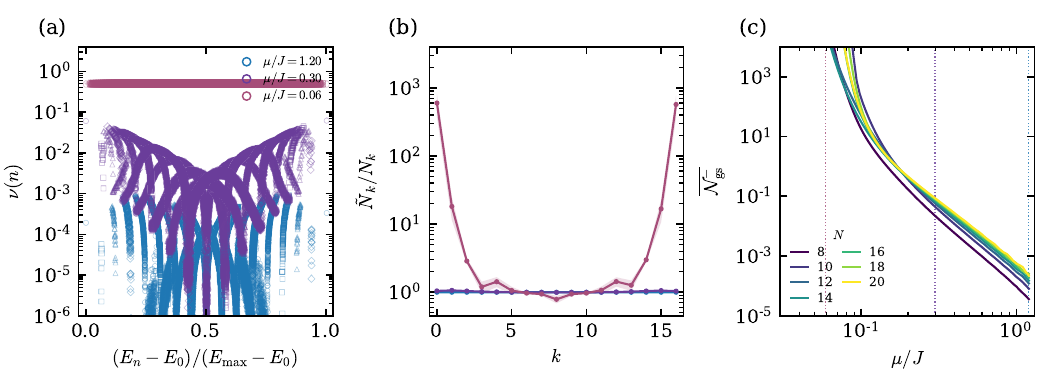}
    \caption{Negativity of the Kirkwood--Dirac cluster weights. (a) Negativity fraction $\nu(n)=\sum_k\max(0,-\mathcal{W}_k)/\sum_k|\mathcal{W}_k|$ per eigenstate versus the normalized excitation energy $(E_n-E_0)/(E_{\max}-E_0)$, for $N=16$ (one disorder realization) at the three couplings of Fig.~\ref{fig:appC_bare_vs_dressed}. At $\mu=1.20$ and $\mu=0.30$ the negativity is small ($\nu\lesssim10^{-3}$ and $\lesssim6\times10^{-2}$, respectively); at $\mu=0.06$ it saturates the ceiling $\nu=1/2$ for the entire spectrum. (b) Ratio $\tilde N_k/N_k$ of the clipped to the exact degeneracy ($N=16$, averaged over $10$ realizations, same three couplings as panel (a)): the exact sum rule of Eq.~\eqref{eq:sum_rules} is preserved to $10^{-3}$ at $\mu=1.20$ and to a few percent at $\mu=0.30$, and is strongly violated at $\mu=0.06$, where the clusters have merged. (c) Disorder-averaged ground-state negativity $\overline{\mathcal{N}^-_{\rm gs}}=\overline{\sum_k\max(0,-\mathcal{W}_k(E_0))}$ versus $\mu$, for $N=8$--$20$ (since $\sum_k\mathcal{W}_k=1$ exactly, this is also the excess positive weight removed by the clipping); it decreases smoothly with $\mu$ and exceeds unity for $\mu\lesssim0.10$. Dotted lines mark the three couplings of panel (a).}
    \label{fig:appC_negativity}
\end{figure}

For visualization and statistics we use the positive normalized weights
\begin{equation}
    W_k(n) = \frac{\max\!\big(0,\,\mathcal{W}_k(n)\big)}{\sum_{k'}\max\!\big(0,\,\mathcal{W}_{k'}(n)\big)},
    \label{eq:W_positive}
\end{equation}
which inherit $\sum_k W_k(n)=1$ but satisfy $\sum_n W_k(n)=\tilde N_k\neq N_k$. This clipping is faithful only if the discarded negative part is small. Figure~\ref{fig:appC_negativity} quantifies when this holds, using the negativity fraction $\nu(n)=\sum_k\max(0,-\mathcal{W}_k(n))/\sum_k|\mathcal{W}_k(n)|\in[0,1/2]$ [panel (a)] and the degeneracy ratio $\tilde N_k/N_k$ [panel (b)]. The answer tracks the cluster structure itself. In the resolved regime the clipping is essentially exact: at $\mu=1.20$ the negativity never exceeds $\nu\sim10^{-3}$ and $\tilde N_k/N_k$ deviates from unity at the $10^{-3}$ level, while at $\mu=0.30$ the negativity stays below $\nu\lesssim6\times10^{-2}$ and the degeneracy distortion is at most a few percent. Below $\mu_c(N)$, where the clusters merge, the quasiprobability itself degrades: the dressing map becomes singular ($\beff\sim1/\mu$ diverges, and the oblique projectors become extremely ill-conditioned), the raw weights grow to $|\mathcal{W}_k|\sim10^5$ with near-perfect cancellations, and $\nu$ saturates its ceiling of $1/2$ across the whole spectrum. This same breakdown of the dressing map, tracked across $N=8$--$20$, is shown in panel (c). In that regime the positive weights $W_k(n)$ remain a well-defined assignment, but they should be read as a heuristic cluster label rather than a faithful proxy for the quasiprobability. All quantitative cluster analysis in the paper (centroids, widths, thermodynamics) is restricted to the resolved window $\mu>0.1$ for $N=16$, where the clipping is quantitatively controlled. The data displayed at $\mu=0.06$ serve only to illustrate the breakdown of the clustering scheme.

We identify $W_k(n)$ as the probability that eigenstate $n$ belongs to cluster $k$ and define the \emph{dominant cluster assignment} $k_*(n)=\operatorname{argmax}_k W_k(n)$. The cluster \emph{dominance}
\begin{equation}
    \mathcal{D}(n) = W_{k_*}(n)
    \label{eq:dominance}
\end{equation}
measures how sharply a state sits in one sector: $\mathcal{D}\approx1$ is hard clustering, while $\mathcal{D}\approx0.5$ signals strong mixing between two sectors. The \emph{effective cluster number} is the inverse participation ratio in $k$-space,
\begin{equation}
    \mathcal{N}_{\mathrm{eff}}(n) = \left(\sum_{k=0}^N W_k(n)^2\right)^{-1},
    \label{eq:neff}
\end{equation}
which counts how many sectors participate: $\mathcal{N}_{\mathrm{eff}}=1$ for a state localized in one sector and $\mathcal{N}_{\mathrm{eff}}=M$ for equal mixing across $M$ sectors. Size averages are defined as $\langle\cdots\rangle_k=\tilde N_k^{-1}\sum_n W_k(n)(\cdots)$, so that the cluster centroid is
\begin{equation}
    \bar E_k = \frac{\sum_n W_k(n)\,E_n}{\sum_n W_k(n)}.
    \label{eq:centroid}
\end{equation}

The centroids that build the towers in the main text are not extracted from the full weighted average of Eq.~\eqref{eq:centroid}, but from its restriction to high-dominance eigenstates. In practice we keep only states with $\mathcal{D}(n)>0.7$ and compute the centroid of that subset. This is robust because the retained states sit deep inside a single cluster, where the soft membership is essentially binary and insensitive to the small negative tails of the quasiprobability; the low-dominance states that straddle cluster boundaries, which carry most of the ambiguity, are simply excluded. The resulting $\bar E_k$ reproduce the peak positions of the spectral functions.

\subsection{Benchmark against the exact adiabatic flow}
\label{app:adiabatic_benchmark}

At $N=12$ the exact projectors $\mathcal{P}_k(\mu)$ of Eq.~\eqref{eq:transported} can be built numerically, so the accuracy of the replacement $U(\mu,\mu_0)\to S(\mu)$ can be tested. Since the transported space
$\mathcal{P}_k(\mu)\mathcal{H}$ is an invariant subspace of $H(\mu)$, it is spanned by
$N_k$ of its eigenvectors, and the flow reduces to assigning each eigenvector a label
$k$. As both the adiabatic flow and the approximate dressing preserve the
$\mathbb{Z}_4$ symmetry, we restrict our analysis to the ground state sector; for
$N=12$ it contains $k=0,4,8,12$ and has dimension $D_0=1+495+495+1=992$ out of $4096$.
We start at $\mu=8$, where clusters are well separated and the bare weights already label every eigenstate
unambiguously, and walk down to $\mu=0.03$ in $4000$ logarithmic steps, diagonalizing
$H(\mu)$ at each step and matching new eigenvectors to old ones by maximal overlap.
We only need the label $k$, not the identity of the individual eigenstates, so a
crossing between two states that carry the same label does not affect the result. As
a consistency check we verify at each step that every label still carries exactly
$N_k$ states, which holds over the whole sweep.

The tracking is robust: sweeping back up to $\mu=8$ returns every eigenstate to the
label it started from, in all $50$ disorder realizations, and halving the step size
changes no label. The adiabatic labels are therefore well defined down to
$\mu=0.03$, so what breaks down at small $\mu$ is the dressing and not the labels it
approximates.

We compare the adiabatic projectors $\mathcal{P}_k(\mu)$ with the dressed ones $\tilde{P}_k(\mu)$ through their trace overlap,
\begin{equation}
    T_{kk'}(\mu)=\frac{1}{D_0}\Tr\!\left[\tilde P_k(\mu)\,\mathcal{P}_{k'}(\mu)\right]
    =\frac{1}{D_0}\sum_{n\,\in\,k'}\mathcal{W}_k(n),
    \label{eq:confusion}
\end{equation}
where the sum runs over the eigenstates that the flow assigns to $k'$. Because both
families of projectors resolve the identity, the sum rules of
Eq.~\eqref{eq:sum_rules} fix the row and column sums of $T$ to $N_k/D_0$ and
$N_{k'}/D_0$ whatever the quality of the dressing. The fraction of the exact cluster
$k'$ that the dressing assigns to $k$ is $M_{kk'}=T_{kk'}D_0/N_{k'}$, equal to the
identity for an exact dressing. As a scalar diagnostic we use
\begin{equation}
    \Delta_{\rm TV}(\mu)=\frac{1}{2}\sum_{k,k'}
    \left|T_{kk'}-\delta_{kk'}\frac{N_k}{D_0}\right| .
    \label{eq:dtv}
\end{equation}
Since the $\mathcal{W}_k(n)$ can be negative and arbitrarily large, $\Delta_{\rm TV}$
is not bounded. We evaluate $T$ for the raw $\mathcal{W}_k(n)$, for the positive
weights $W_k(n)$ that the analysis uses, and for the bare weights
$\bra{E_n}P_k\ket{E_n}$, which are what the dressing has to improve on.

\begin{figure}
    \centering
    \includegraphics[width=\linewidth]{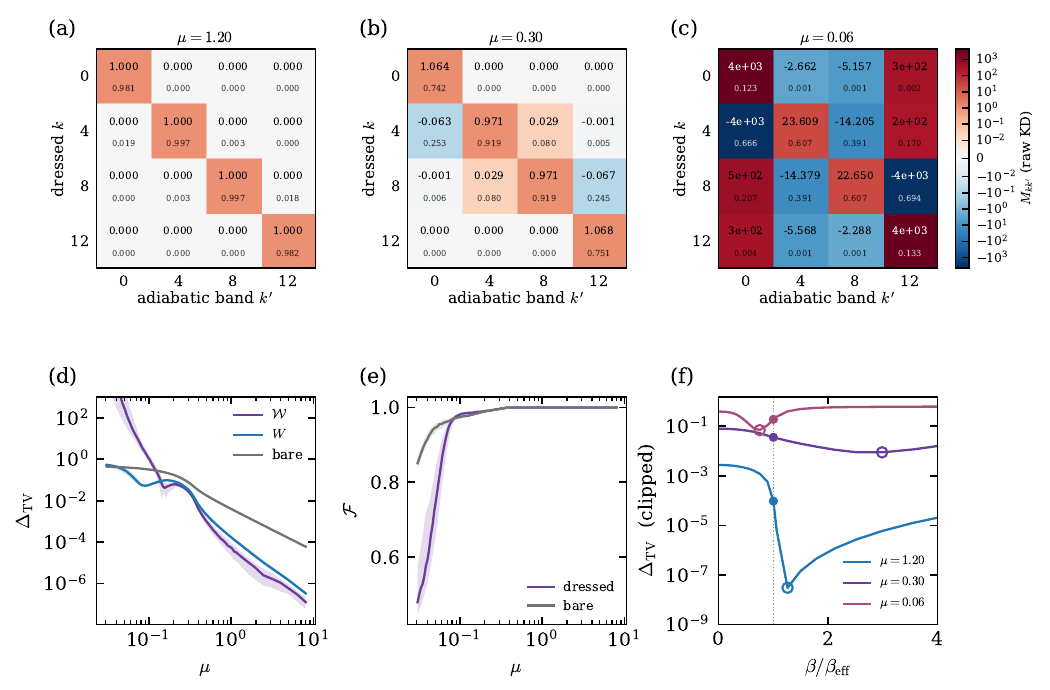}
    \caption{Dressed oblique projectors compared with the exact adiabatic ones at
    $N=12$, in the $\mathbb{Z}_4$ sector of the ground state ($k=0,4,8,12$,
    $D_0=992$). (a)--(c) Confusion matrix built from the fraction $M_{kk'}=T_{kk'}D_0/N_{k'}$ of the exact cluster
    $k'$ that the dressing assigns to $k$, at $\mu=1.20,\,0.30,\,0.06$ for one
    disorder realization. The large number and
    the color give the raw quasiprobability, on a diverging color map, logarithmic
    away from zero. The gray number represents the same confusion matrix but built from the bare size operator, without doing any dressing. (d) Distance
    $\Delta_{\rm TV}$ of Eq.~\eqref{eq:dtv} between $T$ and the exact result.
    (e) Fraction $\mathcal{F}$ of the $992$ eigenstates whose dominant label $k_*(n)$ equals the adiabatic one. (f) $\Delta_{\rm TV}$ of the positive weights versus
    $\beta/\beff$ at the three couplings, with $\beta=\beff$ marked by the vertical
    line and by the filled circles, and the optimum by the open circles. Lines in
    (d)--(f) are medians over $50$ disorder realizations, and the bands in (d) and
    (e) the $16$--$84$ percentiles.}
    \label{fig:appC_benchmark}
\end{figure}

Panels (a)--(c) of Fig.~\ref{fig:appC_benchmark} show $M$ at the three couplings used
throughout the text. At $\mu=1.20$ the dressing is exact to the digits shown, $M_{kk}=1.000$ for the four labels, against $0.981$--$0.997$ for the bare projectors. At $\mu=0.30$ the dressed diagonal reads $1.064,\,0.971,\,0.971,\,1.068$ while the bare diagonal drops significantly to $0.742,\,0.919,\,0.919,\,0.751$. At $\mu=0.06$ the entries reach $\pm4\times10^{3}$ and the dressed labels have lost all meaning. This breakdown is a property of the oblique projector, which can no longer represent the adiabatic flow.

Panel (d) follows $\Delta_{\rm TV}$ over the whole range of $\mu$. At $\mu=1.20$ the dressing improves on the bare projectors by a factor of $29$, from $2.8\times10^{-3}$ to $9.6\times10^{-5}$, and the raw quasiprobability $\mathcal{W}_k$ becomes the worse estimator only below $\mu=0.118$; the clipped weights $W_k$ stay better than the bare ones until $\mu=0.040$. Panel (e) shows the fraction $\mathcal{F}$ of eigenstates whose dominant label $k_*(n)$ matches the adiabatic one. It equals one for both the dressed and the bare weights at large $\mu/J$ and gradually degrades until the dressing becomes worse than the bare assignment at $\mu=0.081$. This crossing falls in the window $\mu\approx0.06$--$0.13$ determined by $\mu^*$ and $\mu_\times$ (Fig.~\ref{fig:appC_vacuum}). The dressed projectors are thus faithful down to the coupling at which the clusters merge, and the restriction to $\mu>0.1$ used for the analysis in the main text lies above all three crossings.

Panel (f) evaluates whether the optimal $\beta$ is $\beff$ by recomputing $\Delta_{\rm TV}$ for different values of $\beta$. At $\mu=1.20$ the optimum is $1.26\beff$, but the classification is already essentially perfect at $\beff$ and the gain is marginal. The only clear gain is at $\mu=0.059$, where $0.75\,\beff$ raises the fraction $\mathcal{F}$ of correct $k_*(n)$ from $0.841$ to $0.967$. We do not know how to compute the optimal $\beta$ without doing the adiabatic calculation first, but moving to it does not change the assignment in the
resolved window, which justifies the use of $\beff$ throughout this work.

Taken together, these tests show that for the system sizes studied both approximations, $U(\mu,\mu_0)\to S(\beff(\mu))$ and $\ket{E_0}\approx\ket{\mathrm{TFD}(\beff)}$, break down only at roughly the coupling at which the clusters merge, so the analysis of the main text is faithful throughout the resolved window.

\section{Green's Functions and Revivals}\label{app:RevivalsGF}
\subsection{Dynamic Observables and Spectral Decomposition}
To probe the dynamical properties of the two coupled SYK model, we rely on the real-time greater Green's function for a generic local or non-local operator $O$. At a finite inverse temperature $\beta = 1/T$, this is defined as the thermal expectation value
\begin{equation}
    G_{O}^{>}(t, \beta) = -i \langle O(t)O(0) \rangle_{\beta} = -\frac{i}{Z(\beta)} \Tr \left[ e^{-\beta H} O(t)O(0) \right].
\end{equation}

The spectral content governing this dynamics is captured by the corresponding spectral function. By inserting a complete set of energy eigenstates $H \ket{E_n} = E_n \ket{E_n}$, we can express the Green's function in frequency space using the Lehmann representation. The spectral function for an arbitrary operator $O$ at finite temperature takes the form
\begin{equation}
    A_{O}(\omega, \beta) = \sum_{m,n} \frac{e^{-\beta E_m} \pm e^{-\beta E_n}}{Z(\beta)} |\bra{E_n}O\ket{E_m}|^2 \delta(\omega - (E_n - E_m)),
\end{equation}
where the sign depends on the fermionic or bosonic statistics of the operator. This general formulation establishes a direct link between the real time dynamics and the microscopic many-body spectrum; the shape of the spectral function is dictated by the energy transitions $(E_n - E_m)$ allowed by the operator $O$, weighted by the thermal Boltzmann factors.

In large-$N$ numerics~\cite{Lantagne2020,FloquetWormhole}, the conformal spectrum is identified through the poles of $A_O(\omega, \beta)$. While finite temperature dynamics involve a vast number of transitions that can smear the spectral peaks into a continuum, the traversable wormhole phase and its associated spectrum are most sharply defined in the low-temperature and low-energy limit. Taking the zero-temperature limit ($\beta \to \infty$), the thermal density matrix projects entirely onto the ground state:
\begin{equation*}
    \frac{e^{-\beta E_m} \pm e^{-\beta E_n}}{Z(\beta)} \xrightarrow{\beta \to \infty} \begin{cases} 1 & \text{if } m=0 \text{ and } n \neq 0 \\ \pm1 & \text{if } n=0 \text{ and } m \neq0\\ 0 & \text{otherwise} \end{cases}
\end{equation*}
The second branch contributes only at $\omega<0$; we retain the $\omega>0$ half of the spectral function throughout.

Consequently, the spectral function collapses to exclusively describe excitations above the ground state
\begin{equation}
    A_{O}^{\beta \to \infty}(\omega) = \sum_{n} |\bra{E_n} O \ket{E_0}|^2 \delta(\omega - (E_n - E_0))
\end{equation}

\subsection{Probing the Holographic Towers}
With the zero-temperature spectral function defined, we can independently access the distinct excitation towers predicted by the semiclassical gravity dual by selecting the appropriate operator $O$.

To probe the matter sector, we use the single-fermion operator $O = \chi_j^a$ with $a \in \{L, R\}$, and define the matter greater Green's function, averaged over all the $N$ Majorana modes, as
\begin{equation}
    G_{ab}^{>}(t) = -\frac{i}{N} \sum_{j=1}^{N} \langle \chi_j^a(t)\chi_j^b(0) \rangle.
\end{equation}
This two-point correlation function is the mathematical object relevant for large-$N$ calculations~\cite{MaldaStanford, MQ}.
The hallmark dynamical signature of the traversable wormhole phase, i.e., the coherent transmission of information between the two boundaries, is quantified precisely by the transmission amplitude \cite{MQ, Lantagne2020}
\begin{equation}
    T_{ab}(t) = 2|G_{ab}^{>}(t)|\,.
\end{equation}
To understand the spectral origin of this transmission, we analyze its associated spectral function $A_{ab}^{\beta \to \infty}(\omega)$. We define the single-fermion excitation over the ground state as $|\chi_j^a\rangle = \sqrt{2}\chi_j^a|E_0\rangle$. Because the chirality operator acts on the state as $\gamma_5 |\chi_j^a\rangle = -\lambda_{0}^2 |\chi_j^a\rangle$, this operator strictly couples the ground state to excited states within the symmetry sectors $\lambda_{1,3} = \pm i\lambda_{0}$. Expanding $|\chi_j^a\rangle = \sum_{\lambda,n} (c_j^a)_n^\lambda |E_n^\lambda\rangle$, the diagonal ($a=b$) spectral function becomes
\begin{equation}
    A_{m}^{\beta \to \infty}(\omega) = \frac{1}{2N}\sum_{j}\sum_{\lambda=\lambda_1, \lambda_3}\sum_{n=1}^{N_\lambda} |(c_j^a)_n^\lambda|^2 \delta(\omega - (E_n^\lambda - E_0))\,.
\end{equation}
In large-$N$ calculations, the discrete poles of this function are identified with states in the conformal matter tower~\cite{MQ, Lantagne2020}. The position of the first peak defines the conformal gap, $E_{\rm gap}^{(m)}\sim \varepsilon/q \sim J(\mu/J)^{\frac{q}{2(q-1)}}=\mu^{2/3}$ (for $q=4$ and $J=1$). In the large-$N$, $\mu/J\to0$ limit, the conformal tower takes the form $E_n^{(m)} = E_{\rm gap}^{(m)}(q n+1)$.

Conversely, the graviton tower associated with boundary reparametrization modes is probed using the two-fermion inter-boundary operator $O_g^j = i\chi_j^L\chi_j^R$. This choice is physically motivated by quench dynamics: time-evolving an initial thermofield double state $|TFD(\beta)\rangle$ under $H(\mu)$ produces collective boundary oscillations that correspond to these graviton modes~\cite{Schuster2025}.
The associated dynamical response is captured by the four-point Green's function
\begin{equation}
    \tilde{G}^{>}(t) = -\frac{i}{N} \sum_{j=1}^{N} \langle O_g^j(t) O_g^j(0)\rangle = \frac{i}{N} \sum_{j=1}^{N} \langle\chi_j^L(t)\chi_j^R(t)\chi_j^L(0)\chi_j^R(0) \rangle\,.
\end{equation}

Applying the operator $O_g$ to the ground state yields the state $\ket{\chi_j^L\chi_j^R} = 2i\chi_j^L\chi_j^R\ket{E_0}$. Unlike the single-fermion operator, $O_g^j$ preserves the symmetry sector of the ground state, exclusively exciting states within the $\lambda = \lambda_{0}$ subspace. Defining the expansion coefficients as $g_{nj} = \langle E_n^{\lambda_{0}} | \chi_j^L\chi_j^R \rangle$, the associated spectral function is
\begin{equation}
    A_g^{\beta \to \infty}(\omega) = \frac{1}{4N}\sum_{j}\sum_{n=1}^{N_{\lambda_{0}}} |g_{nj}|^2 \delta(\omega - (E_n^{\lambda_{0}} - E_0))\,.
\end{equation}
The first peak in this distribution defines the graviton gap, $E_{\rm gap}^{(g)}$, which again scales as $\varepsilon$. The large-$N$ graviton tower is expected to follow $E_n^{(g)}=E_{\rm gap}^{(g)}(2n+1)$. Because the single-fermion and two-fermion operators access different symmetry sectors, their respective energy gaps follow a strict hierarchy. Since the absolute spectral gap of the system is $E_{\rm gap} = E_1 - E_0$, and the first excited state $\ket{E_1}$ typically belongs to the $\lambda_{1}$ sector, we obtain $E_{\rm gap} \le E_{\rm gap}^{(m)} < E_{\rm gap}^{(g)}$.

The $\mathbb{Z}_4$ selection rules also explain, in hindsight, the resonance pattern found when the coupling $\mu(t)$ is driven periodically~\cite{FloquetWormhole}. In that setup the response is read off the energy $\langle H\rangle$ and the drive acts through $\Hint$. Both operators are $\mathbb{Z}_4$-neutral, so they can only induce transitions with $\Delta k\equiv 0\ (\mathrm{mod}\ 4)$. The initial state is thermal, but at low temperature in the wormhole phase it is dominated by the ground state, which sits in the $k=0$ cluster. The drive can then only reach the clusters $k=4,8,\dots$, so it cannot excite the matter tower directly: it can only create matter excitations in \emph{pairs}. A pair on the tower levels $E^{(m)}_n$ and $E^{(m)}_{n^\prime}$ has total size $(2n+1)+(2n^{\prime}+1)$, which is a multiple of $4$ only if $n+n^{\prime}$ is odd. The cheapest such pairs keep one excitation at the bottom of the tower, so the absorption thresholds are $\Omega_M=E^{(m)}_{2M+1}+E^{(m)}_0$ with $M=0,1,2,\dots$ Since the tower is evenly spaced, these thresholds are separated by twice the tower spacing. Thus, the drive responds to only half of the spectrum, not because the other half is suppressed, but because reaching it would require an odd $\Delta k$, which the $\mathbb{Z}_4$ symmetry forbids. The strongest resonance is the graviton, the bound state of the pair channel, which sits just below the first threshold and carries most of the spectral weight. Thermal population of the odd clusters does not change this picture since an occupied matter level can only be moved by $\Delta k=\pm4$, i.e., $E^{(m)}_n\to E^{(m)}_{n\pm2}$. These transitions are Boltzmann-suppressed, and for an evenly spaced tower they all fall on pure multiples of twice the tower spacing, with no offset, so they cannot be confused with the shifted pair comb $\Omega_M$.

\subsection{Gravitational Signals vs. Finite-\texorpdfstring{$N$}{N} Effects}
The spectral structure derived above provides the microscopic framework to understand the time-domain transmission amplitudes $T_{ab}(t)$. In the holographic (large-$N$) wormhole phase, the zero temperature spectral function $A_{aa}^{\beta \to \infty}(\omega)$ is not a featureless continuum but exhibits evenly spaced poles. At finite $N$, exact diagonalization resolves the precursors of these poles as peaks whose origin, as shown in the main text, is the spectral clustering.

In the time domain, this regular spacing of dynamical resonances acts effectively as a discrete Fourier series, driving the coherent reassembly of the initial perturbation and producing the revivals observed in $T_{ab}(t)$. However, periodicity is spoiled by the finite cluster widths as the states within a cluster acquire different phases and fail to realign exactly after each transmission cycle. This microscopic dephasing drives the gradual loss of revival fidelity over time, smoothly erasing the transmission amplitude as the system thermalizes. At finite temperature the same picture accounts for the small but nonzero width of the resonances reported in Ref.~\cite{Qi2020}. The spectral function receives contributions $\delta(\omega-(E_n-E_m))$ with $m\neq0$, and any residual non-uniformity in the cluster spacing turns each resonance into a narrow bundle of $\delta$-peaks rather than a single one.

To firmly establish the gravitational origin of these revivals, two key dynamical signatures must be present. First, the transmission must exhibit an alternating oscillatory behavior, where $T_{LL}(t)$ and $T_{LR}(t)$ exchange their local maxima and minima, reflecting the physical traversal of the signal between boundaries. Second, the revival frequency, governed by $E_{\rm gap}^{(m)}$, must follow the conformal scaling $\varepsilon$ predicted by the nearly-AdS$_2$ holographic dual. For the $q=4$ model, this requires the frequency to scale as $\varepsilon=\mu^{2/3}$. At finite $N$, this wormhole-like behavior only emerges in an intermediate coupling regime. At very large $\mu/J$, the SYK interactions become negligible compared to $\Hint$, the spectral spacing becomes trivial (harmonic), and the gravitational description does not apply.

As $\mu$ is further lowered towards zero, a microscopic collision of scales occurs: the conformal gap shrinks according to the $\mu^{2/3}$ scaling, as shown in the main text, but the width of the peaks of the spectral functions does not compress at the same rate. Eventually, this spread overpowers the conformal gap, the clusters severely overlap, and the surviving spectral resonances dissolve into an effectively continuous random matrix theory (RMT) profile. While revivals in this deep infrared regime were reported in Ref.~\cite{Lantagne2020}, they only appear for the smallest values of $N$ and do not have a gravitational origin. We conjecture that the breakdown of the revivals at small $\mu$ is the microscopic manifestation of the wormhole collapsing into a disconnected, black-hole-like phase. This is consistent with the backreaction interpretation conjectured in the main text. Energetically, the traversable wormhole is stabilized by the coupling $\Hint$, which provides a binding energy proportional to $\mu N$. Probing the dynamical transmission via the Green's function effectively injects a single Majorana fermion into the system, introducing a $\mathcal{O}(1)$ energy perturbation. In the strict large-$N$ limit, this single Majorana excitation is entirely negligible compared to the $\mathcal{O}(\mu N)$ background, allowing the signal to traverse without altering the geometry ($G_N \sim 1/N \to 0$). However, at finite $N$ and small $\mu$, the $\mathcal{O}(1)$ energetic ``kick'' of the probe particle becomes comparable to the total stabilization energy. From the gravitational point of view, the injected particle heavily backreacts on the finite-$N$ background, irreversibly destroying the phase coherence required to transmit information. Consequently, observing traversable wormhole-like dynamics at finite $N$ strictly requires an intermediate coupling window: $\mu$ must be small enough to suppress the trivial harmonic behavior in favor of the emergent SYK conformal symmetry, yet large enough to stabilize the geometry against this probe-induced backreaction.

\end{document}